\documentclass[superscriptaddress,secnumarabic,
amssymb,amsmath,nobibnotes,aps,prd,showpacs,nofootinbib]{revtex4}%
\usepackage{graphicx}
\usepackage{epsf}
\usepackage{bm}
\usepackage{amsmath}
\usepackage{amsfonts}
\usepackage{amssymb}
\usepackage{epstopdf}
\usepackage{natbib}
\usepackage{color}
\usepackage{float}
\providecommand{\U}[1]{\protect\rule{.1in}{.1in}}
\newcommand{\be}{\begin{equation}}
\newcommand{\ee}{\end{equation}}

\newcommand{\mincir}{\raise
-3.truept\hbox{\rlap{\hbox{$\sim$}}\raise4.truept\hbox{$<$}\ }}
\newcommand{\magcir}{\raise
-3.truept\hbox{\rlap{\hbox{$\sim$}}\raise4.truept\hbox{$>$}\ }}

\begin{document}

\title{Understanding the phenomenology of interacting dark energy scenarios and their theoretical bounds}

\author{Supriya Pan}
\email{supriya.maths@presiuniv.ac.in}
\affiliation{Department of Mathematics, Presidency University, 86/1 College Street, 
Kolkata 700073, India.}

\author{Jaume de Haro}
\email{jaime.haro@upc.edu}
\affiliation{Departament de Matem\' {a}tica Aplicada, Universitat Polit\`{e}cnica de Catalunya, Diagonal 647, 08028 Barcelona, Spain.}

\author{Weiqiang Yang}
\email{d11102004@163.com}
\affiliation{Department of Physics, Liaoning Normal University, Dalian, 116029, P. R.  China.}

\author{Jaume Amor\'{o}s}
\email{jaume.amoros@upc.edu}
\affiliation{Departament de Matem\' {a}tica Aplicada, Universitat Polit\`{e}cnica de Catalunya, Diagonal 647, 08028 Barcelona, Spain.}

\begin{abstract}
Non-gravitational interaction between dark matter and dark energy has been considered in a spatially flat Friedmann-Lema\^{i}tre-Robertson-Walker (FLRW) universe. The interaction rate is assumed to be linear in the energy densities of dark matter and dark energy and it is also proportional to the Hubble rate of the FLRW universe. This kind of interaction model leads to an autonomous linear dynamical system, and depending on the coupling parameters, could be solved analytically by calculating the  exponential of the matrix, defining the system. We show here that such interaction rate has a very deep connection with some well known cosmological theories. We then investigate the theoretical bounds on the coupling parameters of the interaction rate in order that the energy densities of the dark sector remain positive throughout the evolution of the universe and asymptotically converge to zero at very late times. Our analyses also point out that such linear interacting model may encounter with finite time future singularities depending on the coupling parameters as well as  the dark energy state parameter.  

\end{abstract}
%----------------------------------------------------------------------------
%---------------------------------------------------------------------------- 
\maketitle

%%%%%%%%%%%%%%%%%%%%%%%%%%%%%%%%%%%%%%%%%%%%%%%%%%%%%%%%%%%%%%%%%%%%%%%%%%%%%%
%%%%%%%%%%%%%%%%%%%%%%%%%%%%%%%%%%%%%%%%%%%%%%%%%%%%%%%%%%%%%%%%%%%%%%%%%%%%%%

\section{Introduction}

According to recent observations \cite{Aghanim:2018eyx}, our Universe is currently expanding with an acceleration and this accelerating phase cannot be described by normal matter within the context of General theory of Relativity (GR). A possible approach to explain this accelerating phase of the universe within this context, i.e. within GR, is to consider some exotic matter (characterized by negative pressure), dubbed as dark energy (DE). This dark energy occupies  around 68\% of the total energy budget of our Universe, and this is the largest sector of our Universe \cite{Aghanim:2018eyx}. However, the origin, nature and dynamics of DE are absolutely unknown even after a series of observational missions running since last twenty years. The second largest component of our Universe is dark matter (DM) which takes nearly 28\% of the total energy density of the universe, and similar to the DE sector, this sector is also not very well understood.  Thus, the dynamics of our Universe is mainly driven by DE and DM, the understanding of which has been the central issue for modern cosmology at present.

In order to understand the dynamics of our Universe, mainly the dynamics of the dark sector, usually two different approaches are considered. The first approach is very simple in which DM and DE are independently conserved, that means both DE and DM enjoy their independent evolution. The second approach is a bit complicated but offers a wider picture (compared to the former proposal) where DM and DE might be interacting non-gravitationally with each other. Observations from many sources have already reported that a simplest cosmological model, namely the $\Lambda$-cold dark matter ($\Lambda$CDM), where $\Lambda$ and CDM are independently conserved, is an excellent cosmological model explaining the late accelerating phase. However, in the present work, we are interested in the second approach for several reasons that we are going to describe below. If we consider the $\Lambda$CDM picture of the universe, then we are unable to explain a biggest mystery of the universe, namely, the cosmological constant problem \cite{Weinberg:1988cp}.     
The theory of interaction may play a very crucial role to  offer a satisfactory problem to the cosmological constant problem. In \cite{Wetterich:1994bg} the author showed that a coupled system between gravity and a scalar field with exponential potential could give rise to a cosmological constant term that becomes time-dependent. When the  DE era began, a new problem in the name of cosmic coincidence problem appeared. The interaction in the dark sector again played a crucial role to explain this phenomenon \cite{Amendola:1999er,Amendola:2003eq,Cai:2004dk,Pavon:2005yx,delCampo:2008sr,delCampo:2008jx}. The explanations for both cosmological constant and coincidence problems influenced the scientific community to investigate interacting cosmological models, and consequently, the models in this class soon got massive attention due to having their far reaching activities \cite{Billyard:2000bh,Barrow:2006hia,Amendola:2006dg,He:2008tn,Valiviita:2008iv,Gavela:2009cy,Majerotto:2009np,Clemson:2011an,Pan:2013rha,Yang:2014gza,yang:2014vza,Nunes:2014qoa,Salvatelli:2014zta,Pan:2012ki,Shahalam:2015sja,Tamanini:2015iia,Nunes:2016dlj,Caprini:2016qxs,Kumar:2016zpg,Yang:2016evp,Pan:2016ngu,vandeBruck:2016hpz,Erdem:2016hqw,Sharov:2017iue,Kumar:2017dnp,Shahalam:2017fqt,Cai:2017yww,DiValentino:2017iww,Yang:2017yme,Mifsud:2017fsy,Yang:2017ccc,Yang:2017zjs,Pan:2017ent,Yang:2018xlt,Yang:2018pej,Yang:2018ubt,vonMarttens:2018iav,Yang:2018qec,Paliathanasis:2019hbi,Pan:2019jqh,Yang:2019bpr,Yang:2019vni,Yang:2019uzo,Pan:2019gop,Barrow:2019jlm,Yang:2019uog,DiValentino:2019jae,vonMarttens:2019ixw,Papagiannopoulos:2019kar} (see also two review articles on interacting DE models \cite{Bolotin:2013jpa,Wang:2016lxa}). In particular, an interaction or coupling in the dark sector may naturally push  a quintessence DE, characterized by its equation of state, $w_{DE} = p_{DE}/\rho_{DE} > -1$ to enter the phantom DE phase, $w_{DE} = p_{DE}/\rho_{DE} < -1$ \cite{Wang:2005jx,Sadjadi:2006qb,Pan:2014afa}. Here, $p_{DE}$ and $\rho_{DE}$ are respectively the pressure and energy density of DE. The remarkable point with such phantom crossing is that the crossing of $w_{DE} =  -1$ actually needs some scalar field models with negative kinetic term which automatically invites instabilities both at classical and quantum levels. But, in interacting models with the choice of suitable interaction between DE and DM, the instability problem can be relieved. Later, in the beginning of the year 2016, other observational evidences reported that the estimation of the Hubble constant, $H_{0}$, as well as, the amplitude of the matter power spectrum $\sigma_8$ (within the minimal $\Lambda$CDM cosmology) return different values  in different observational missions that are many sigmas apart from one another. Surprisingly, DE-DM interaction again proved its potential nature by offering a possible solution to the $H_0$ tension \cite{Kumar:2017dnp,DiValentino:2017iww,Yang:2018euj,Yang:2018uae,Kumar:2019wfs,DiValentino:2019ffd} and the $\sigma_8$ tension \cite{Kumar:2019wfs,vandeBruck:2017idm,Pourtsidou:2016ico,An:2017crg}. Certainly, based on the aforementioned limitations of the  non-interacting cosmologies and the solutions coming from the interacting models, it is clear that interacting cosmologies should be investigated more elaborately in light of the above issues.

However, the first and probably the most important question in the context of interaction cosmologies is related to the energy exchange rate  between the dark sectors, that means the interaction function.  Since there is no globally accepted theory yet that could justify the choice of the interaction function, hence, to start with it is assumed that the interaction function maybe involve the energy densities of the dark components ranging from the simple linear to the complicated nonlinear ones. Due to such diverse choices of the interaction function one could explore a cluster of interesting possibilities as a result.  However, as we will show in this work, while dealing with any interaction function, we have to be very careful because depending on the coupling parameter of interaction rate which quantifies the interaction rate in every aspect (whether the interaction rate is mild or not and the direction of energy flow it allows), the associated cosmological parameters could be unrealistic. Thus, in the present work, we show how any interaction rate can be treated leading to physically viable cosmological scenarios. 
We start with a simple interaction model which is linearly related to the energy densities of DM and DE  as well as the Hubble rate of the Friedmann-Lema\^{i}tre-Robertson-Walker universe and then discuss the theoretical bounds on this scenario and the physical consequences. The same analysis can be done for any interaction model irrespective of its linear or nonlinear functional form and consequently the bounds on such scenarios can also be imposed. The work has been organized in the following way: In section \ref{sec-2}, we discuss the field equations of any  interaction model and then introduce our model and provide its justifications. In the next section  \ref{sec-dyn-analysis}, we perform a dynamical system analysis of the original interaction model and its sub-cases and give the bounds on the coupling parameters for realistic interaction scenarios. 
In Section \ref{sec-singularities}, we deal with future singularities, showing that for some values of the coupling parameters and the dark energy equation of state, our universe evolves to a BIG RIP singularity. 
Finally, in section \ref{sec-conclu}, we conclude the main findings of the present article.

\section{Interacting dark energy: Model and Justification}
\label{sec-2}

In the large scale, our Universe is almost homogeneous and isotropic. The geometric configuration of this homogeneous and isotropic universe is  characterized by the Friedmann-Lema\^{i}tre-Robertson-Walker (FLRW) line element

\begin{equation*}
ds^{2}=-dt^{2}+a^{2}(t)\left[ \frac{dr^{2}}{1-Kr^2}+r^{2}(d\theta ^{2}+\sin ^{2}\theta
d\phi ^{2})\right] ,
\end{equation*}%
where $(t, r, \theta, \phi)$ are the co-moving coordinates; $a(t)$ is the expansion scale factor of the universe; $K$ is the curvature scalar of the universe which represents a closed, open  and a flat universe, respectively for  its three distinct values, namely, $K = 1$, $-1$ and $0$. Concerning the gravitational sector of our Universe, we assume that it is described by GR and the matter distribution is minimally coupled to gravity. Precisely, we consider that the total energy density of our
Universe is given by, $\rho_{tot} = \rho_r +\rho_b +\rho_c +\rho_x$, where $\rho_i$ is the energy density of the $i$-th fluid in which 
$i= r, b, c, x$, respectively stands for radiation, baryons, pressure-less DM (also known as cold DM, abbreviated as CDM) and DE. The total pressure contributed due to the above components is therefore given by, $p_{tot} = p_r + p_b +p_c +p_x$, where the notations follow same argument as described above.  
Lastly, we consider that the dark fluids of the universe namely CDM and DE are interacting non-gravitationally with each other, which means that there is a continuous flow of energy and momentum between these sectors. Since other components do not interact with each other, hence, they obey their own conservation equations.    

Therefore, focusing on the interacting dark sector, one can write down the conservation equations of CDM and DE as 
  
\begin{eqnarray}\label{dark}
\left\{\begin{array}{ccc}
\dot{\rho} _{c}+3{H}\rho _{c} &=&-Q\\
\dot{\rho}_{x}+3{H}(1+w_{x})\rho _{x} &=&Q,\end{array}\right.
\end{eqnarray}
where $-1\leq w_x = p_x/\rho_x<-1/3$ (non-phantom fluid), is the constant equation-of-state (EoS) parameter  of the DE fluid and the quantity $Q$ appearing in the above two equations, is the energy transfer function that determines the rate of  energy flow between the fluids as well as the direction of energy flow depending on its sign.  

The conservation equations for the non-interacting radiation and baryonic matter are respectively
\begin{eqnarray}
\dot{\rho}_r+4H\rho_r = 0, \quad \mbox{and}\quad \dot{\rho}_b+3H\rho_b=0, \end{eqnarray}
and introducing the time variable $N\equiv -\ln(1+z)=\ln(a/a_0)$ which is only  the number of e-folds starting at the present time, taking into account that $\dot{N}=H$, one gets
\begin{eqnarray}
{\rho}'_r+4\rho_r=0, \quad \mbox{and}\quad {\rho}'_b+3\rho_b=0, 
\end{eqnarray}
where the prime is the derivative with respect the time $N$. The solutions have the simple form
\begin{eqnarray}
\rho_r=\rho_{r,eq}e^{-4(N-N_{eq})},\quad \mbox{and}\quad \rho_b=\rho_{b,0}e^{-3N},
\end{eqnarray}
where $\rho_{r,eq}$ is the value of the radiation energy density at the matter-radiation equality 
and $\rho_{b,0}$ is the present value of  baryonic energy density.

Once again, in terms of the time $N$ the system of equations in  (\ref{dark}) becomes
\begin{eqnarray}\label{dark1}
\left\{\begin{array}{ccc}
{\rho}'_{c}+3\rho _{c} &=&-\frac{Q}{H}\\
{\rho}'_{x}+3(1+w_{x})\rho _{x} &=&\frac{Q}{H}.\end{array}\right.
\end{eqnarray}
Note that, the Hubble parameter $H$ can be found from the following equation
\begin{eqnarray}
H^2 (N) = \frac{1}{3M_{pl}^2}\left(\rho_c+\rho_x+ \rho_{r,eq}e^{-4(N-N_{eq})}+ \rho_{b,0}e^{-3N}\right) -\frac{K}{a_0^2e^{2N}}  ,
\end{eqnarray}
where $a_0$ is the present value of the scale factor.
Hence, once the evolution of $\rho_c$ and $\rho_x$ are determined 
either analytically or numerically for some given interaction rate $Q$, 
the expansion rate of the  universe  can be determined and the modified cosmological parameters can be studied in terms of their evolution with time.   
Thus, as one can realize, the expansion rate of the universe is highly influenced by the interaction function.  That means the expansion rate is dependent on the choice of $Q$.

Almost in all works, the  choice of the interaction function  is motivated from the phenomenological ground. If we look at the conservation equations in (\ref{dark1}), one can realize that the interaction rate might be the function of the energy densities of the dark sectors, namely, $\rho_c$ and $\rho_x$. Therefore, using that ground as a basis, an infinite number of interaction rates can be produced by hand and can be investigated.  However, some recent investigations show that one could justify the choice of the interaction rates from some field theoretical point of view \cite{ vandeBruck:2015ida,Boehmer:2015kta,Boehmer:2015sha,Gleyzes:2015pma,Amico:2016qft,Pan:2020zza}. In particular, in \cite{Pan:2020zza}, the authors have explicitly shown that some very well known linear and non-linear interaction models can be deduced from scalar field theory. {\it Interestingly, in this work  we shall show that the interaction rates can also be motivated from other well known cosmological backgrounds.} We begin our discussions with a simplest interaction model of the form 

\begin{eqnarray}\label{modelA}
 Q= 3H(\mu\rho_c+\nu\rho_x),
\end{eqnarray}
where $\mu$ and $\nu$ are dimensionless coupling parameters. As one can see from (\ref{modelA}), this interaction rate recovers some well known interaction rates as special cases. For instance, one can recover $Q  = 3 H \mu \rho_c$ under the assumption of $\nu =0$. Similarly, the model $Q = 3 H \nu \rho_x$ is obtained when $\mu =0$. Lastly, for $\mu = \nu = \lambda$, one recovers $Q = 3 H \lambda (\rho_c +\rho_x)$. One could further notice that for $\nu = -\mu$, the interaction function becomes $Q = 3 H \mu (\rho_c - \rho_x),$ which has a sign changeable feature. 
Let us now try to justify the interaction rate (\ref{modelA}) using the available theories in the next paragraphs.

One of the possible justifications of the interaction rate (\ref{modelA}) may appear using  the Teleparallel Gravity (TG), based in the Weitzenb\"ock spacetime \cite{W},
 which is equivalent to General Relativity (GR) \cite{Andrade,Li} (see also \cite{Unzicker} which translates the early papers of Einstein about teleparalellism). Effectively, in TG  to obtain the field equations  the scalar torsion quantity ${\mathcal T}$ is used, which for the flat FLRW space-time is given by
${\mathcal T}=-6H^2$ \cite{Haro}. Additionally, in TG  the total stress tensor satisfies the conservation equation $\nabla_{\alpha}T^{\alpha}_{\beta}=0$ (see for instance \cite{Andrade}), where $\nabla$ denotes the usual
Levi-Civita derivative. Therefore, in this framework  and following \cite{Valiviita:2008iv} we consider the conservation equations in presence of an interaction as $\nabla_{\alpha}T^{\alpha}_{\beta, A}=Q_{\beta, A}$
with $A=c,x$, and 
\begin{eqnarray}
Q_{\beta, c}=-Q_{\beta, x}=\sqrt{{\mathcal T}/6}\left(\bar{\mu}T^{\alpha}_{\alpha,c}u_{\beta,c}+\bar{\nu}T^{\alpha}_{\alpha,x}u_{\beta,x}\right)
\end{eqnarray}
where for a perfect fluid $T^{\alpha}_{\beta, A}=P_A\delta^{\alpha}_{\beta}+(\rho_A+P_A)u^{\alpha}_Au_{\beta,A}$ in which $u^{\alpha}_A=\frac{dx^{\alpha}}{\sqrt{-ds^2}}$ being the four-velocity of the fluid.

Then, since at the background level 
\begin{eqnarray}
Q=Q_{0,c}=-Q_{0,x}=\frac{H}{\sqrt{6}}\left( \bar{\mu}\rho_c+\bar{\nu}(3w_x-1)\rho_x
\right),
\end{eqnarray}
and $\nabla_{\alpha}T^{\alpha}_{\beta, A}=-\dot{\rho}_A-3H(\rho_A+P_A)$, 
to  recover our energy density transfer and  the system (\ref{dark1}), one only has to choose $\mu=\frac{\bar{\mu}}{3\sqrt{6}}$ and $\nu=\frac{(3w_x-1)\bar{\nu}}{3\sqrt{6}}$.

On the other hand, following the notation of \cite{Valiviita:2008iv} the perturbed four-velocity of the B-fluid is given by $u^{\alpha}_B=(1-\phi,\frac{1}{a}\partial ^i v_B)$, where $\phi$ is the Newtonian potential, $v_B$ is the particular velocity and $B=r,b,c,x$. This means that at the background level the four-velocity, in the co-moving system, becomes $u^{\alpha}_B=(1,0,0,0)$ for any fluid which is the consequence of the isotropy at the background level. Therefore, at the background level the scalar
$u_{\alpha,B}T^{\alpha}_{\beta, B}u^{\beta}$ becomes $\rho_B$, and one can also choose the energy transfer function 
\begin{eqnarray}
Q_{\beta, c}=-Q_{\beta, x}=3\sqrt{{\mathcal T}/6}\left({\mu}u_{\alpha,c}T^{\alpha}_{\gamma,c}u^{\gamma}_cu_{\beta,c}+
{\nu}u_{\alpha,x}T^{\alpha}_{\gamma,x}u^{\gamma}_xu_{\beta,x}\right).
\end{eqnarray}

Following this way, and now working in the framework of GR,  we can consider, for example, the scalar $\nabla_{\alpha} u^{\alpha}_c$ which for the flat FLRW metric leads to $3H$, and thus one may consider
\begin{eqnarray}\label{A}
Q_{\beta, c}=-Q_{\beta, x}=\nabla_{\eta} u^{\eta}_c\left({\mu}u_{\alpha,c}T^{\alpha}_{\gamma,c}u^{\gamma}_cu_{\beta,c}+
{\nu}u_{\alpha,x}T^{\alpha}_{\gamma,x}u^{\gamma}_xu_{\beta,x}\right),
\end{eqnarray}
or
\begin{eqnarray}\label{B}
Q_{\beta, c}=-Q_{\beta, x}=\frac{1}{3} \nabla_{\eta} u^{\eta}_c   \left(\bar{\mu}T^{\alpha}_{\alpha,c}u_{\beta,c}+\bar{\nu}T^{\alpha}_{\alpha,x}u_{\beta,x}\right).
\end{eqnarray}

In addition, in GR one could use the Carminati-McLenaghan invariants \cite{cm}
\begin{eqnarray}
 {\mathcal R}_2\equiv \frac{1}{4}{\mathcal R}_{\mu}^{\nu}{\mathcal R}_{\nu}^{\mu}, \quad \mbox{ and } \quad 
 {\mathcal R}_3\equiv -\frac{1}{8}{\mathcal R}_{\mu}^{\nu}{\mathcal R}_{\gamma}^{\mu}{\mathcal R}^{\gamma}_{\nu},
\end{eqnarray}
whose values for the synchronous co-moving observers 
in the flat FLRW space-time are given by
${\mathcal R}_2=\frac{3}{4}\dot{H}^2$ and ${\mathcal R}_3=-\frac{3}{8}\dot{H}^3$. Thus, in such coordinates, one finds that, $\dot{H}=-2\frac{{\mathcal R}_3}{{\mathcal R}_2}$, and consequently, the scalar curvature, namely $R$, takes the relation $H^2= \frac{{\mathcal R}_3}{{\mathcal R}_2}+\frac{R}{12}$. Following this, one can replace $\nabla_{\eta} u^{\eta}_c$
by the following scalar a
\begin{eqnarray}
 {\mathcal R}\equiv 3\sqrt{\frac{{\mathcal R}_3}{{\mathcal R}_2}+\frac{R}{12}},
\end{eqnarray}
in the eqns. (\ref{A}) and (\ref{B}).  
Finally, going beyond GR one can use a mimetic gravity, based on a mimetic field, namely $\varphi$, satisfying $\nabla_{\alpha}\varphi \nabla^{\alpha}\varphi=-1$
(see \cite{mukhanov,mukhanov1} for a detailed description of this theory). Thus,  in mimetic gravity  the scalar $\chi=-\nabla^{\alpha}\nabla_{\alpha}\varphi$ in the flat FLRW space-time becomes $3H$. Thus, one can see that the present interaction model can be justified using various well known cosmological theories.

Let us note that although the interaction rate (\ref{modelA}) is already existing in the list of some well known linear interaction rates, however, below we shall show why we should reconsider the same model. 

To conclude this section we note that the system of first order differential equations given in  (\ref{dark1}) depicts a two dimensional autonomous dynamical system, which could be solved once one has the values of the corresponding energy densities  at  present time.

\section{Dynamical System Analysis}
\label{sec-dyn-analysis}

In this section we shall perform a detailed dynamical analysis of the interaction model (\ref{modelA}) as well as the simple cases of (\ref{modelA})
aiming to provide with the bounds on the coupling parameters of this interaction model. Let's start our analysis with the general interaction model (\ref{modelA}) for which the dynamical system becomes
 \begin{eqnarray}\label{system}
 X'=AX
 \end{eqnarray}
 where prime, as already mentioned earlier, denotes the differentiation with respect to $N$ and 
 \begin{eqnarray}
 A= \left( \begin{array}{cc}
 -3(\mu+1) & -3\nu\\
 3\mu & -3(1+w_x-\nu)\end{array}\right), \quad \quad  X=\left(  \begin{array}{c} \rho_c\\
 \rho_x\end{array}\right).
 \end{eqnarray}
 
 Since we are dealing with a non-degenerate  linear dynamical system, the only fixed point is the origin, and its stability is deduced using the Trace-Determinant criterion.
 Physically, we want that the origin is an attractor, because in the contrary case we will have future singular solutions,  so we have to impose the condition  
 $\mbox{Tr}\; A=-3(2+\mu-\nu+w_x)<0$ and $\mbox{Det}\; A= 9(\mu+1)(1+w_x-\nu)+9\mu\nu>0$. These lead to the allowed region in the plane of parameters $(\mu, \nu)$ determined by the linear inequalities 
 \begin{eqnarray}\label{region}
 \left\{\begin{array}{ccc}
 \nu &<& 2+\mu+w_x,\\
 \nu &<&  (1+w_x)(\mu+1),\end{array}\right.
  \end{eqnarray}
 which involves the EoS of DE along with the coupling parameters. In particular, if 
the DE is assumed to be the vacuum energy characterized by  $w_x=-1$, the above domain becomes  
 \begin{eqnarray}D=\left\{(\mu,\nu): \mu>\nu-1, \nu<0\right\}.
 \end{eqnarray}
 
 However, the condition that the origin is an attractor is not enough, because if the origin is an attractor focus, the orbits will round around $(0,0)$, and thus, the energy densities will be negative, which has no physical sense. For this reason we have to demand that the discriminant $\Delta= (\mbox{Tr} A)^2-4\mbox{Det}\; A$ has to be positive, that means, 
 \begin{eqnarray}
\Delta=9\left( (w_x-\mu-\nu)^2-4\mu\nu\right)>0.
 \end{eqnarray}
 
 In addition, if we want that the energy densities must be positive all the time, then we also need to demand that the eigenvectors of $A$, namely 
 $\vec{v}_{\pm}=(v_{\pm,1},v_{\pm, 2})$,
 have to stay in the first quadrant, i.e., they have to satisfy $v_{\pm,1}v_{\pm, 2}\geq 0$. Effectively, since the system is autonomous, the orbits never cross, then with this condition, the orbits  $\{e^{\lambda_+N}\vec{v}_+\}_{N\in  \mathbb{R}}$ and $\{e^{\lambda_-N}\vec{v}_-\}_{N\in \mathbb{R}}$ define a sector in the first quadrant, and all solution with initial conditions in this sector defines an orbit inside it.

 Denoting the eigenvalues of the matrix $A$ by 
 %\begin{eqnarray}
 $\lambda_{\pm}=\left(\mbox{Tr}\; A \pm \sqrt{\Delta} \right)/2$,
 %\end{eqnarray}
 the corresponding eigenvectors are given by as follows: 
 
 \begin{enumerate}
 \item For $\nu\not=0$:
 \begin{eqnarray}
 \vec{v}_{\pm}=\left(1,-\frac{\mu+1+\lambda_{\pm}/3}{\nu}\right),
 \end{eqnarray}
 and thus, the condition $v_{\pm,1}v_{\pm, 2}\geq 0$, becomes  
 \begin{eqnarray}
 \label{eigenvector}
 \frac{\mu+1+\lambda_{\pm}/3}{\nu}\leq 0.\end{eqnarray}
 
 Then, to have positive energy densities all the time, the initial condition $(\rho_{c,0},\rho_{x,0})$ has to satisfy
 \begin{eqnarray}\label{energydensities0}
 \mbox{min}\left( -\frac{\mu+1+\lambda_{+}/3}{\nu}, -\frac{\mu+1+\lambda_{-}/3}{\nu}   \right)\leq \frac{\rho_{0,x}}{\rho_{0,c}}\leq
 \mbox{max}\left( -\frac{\mu+1+\lambda_{+}/3}{\nu}, -\frac{\mu+1+\lambda_{-}/3}{\nu}   \right). \end{eqnarray}

 \item For $\nu=0$: The eigenvalues are $\lambda_+=-3(1+w_x)$ and $\lambda_-=-3(\mu+1)$, and the corresponding eigenvectors are given by
 \begin{eqnarray}
\vec{v}_+=(0,1),\quad \vec{v}_-=\left(1,\frac{\mu}{w_x-\mu}\right),
 \end{eqnarray}
 and thus, the condition $v_{\pm,1}v_{\pm, 2}\geq 0$, becomes 
 $ w_x\leq \mu\leq 0$.

 Then, the condition to have positive energy densities all the time is
 \begin{eqnarray}\label{energydensities1}
 \frac{\rho_{x,0}}{\rho_{c,0}}\geq  \frac{\mu}{w_x-\mu}. \end{eqnarray}
 
 Taking into account that  $\frac{\rho_{x,0}}{\rho_{c,0}}= \frac{\Omega_{x,0}}{\Omega_{c,0}}$, and as we will see, if one disregards the energy of the radiation at the present
 time one has $\Omega_{c,0}\cong 0.262$ and $\Omega_{x,0}\cong 0.69$, hence, the condition (\ref{energydensities1}) becomes $\mu \geq 0.72 w_x$, which means that the parameter $\mu$ is constrained to satisfy 
 \begin{eqnarray}\label{mu}
 0.72 w_x\leq \mu\leq 0.
 \end{eqnarray}
  
  \end{enumerate}

 On the other hand,  to know the value of the effective EoS parameter and thus, to know if the Universe accelerates or decelerates, we need to calculate explicitly the solutions of 
 (\ref{system}) which is given by $X(N)=e^{AN}X_0$, $X_0$ being the current value of $X$ with 
 \begin{eqnarray}
 e^{AN}=B \left( \begin{array}{cc} e^{\lambda_+N}&0\\
 0& e^{\lambda_- N}\end{array}  \right)  B^{-1},
 \end{eqnarray}
 where, $\lambda_{\pm}$ are once again the eigenvalues and the matrix $B$ is set up with the eigenvectors of $A$,  $\vec{v}_{\pm}$, i.e.,
 \begin{eqnarray}
 B=\left(\begin{array}{cc}
 v_{+,1}& v_{-,1}\\
 v_{+,2}& v_{-,2}\end{array}  \right).
 \end{eqnarray}

 Thus, when $\nu\not=0$, since $\mbox{Det} B=\frac{\sqrt{\Delta}}{3\nu}$, we will have 
 \begin{eqnarray}
 B=\left(\begin{array}{cc}
 1 &  1\\
 v_{+,2}& v_{-,2}\end{array}  \right),  \quad B^{-1}=\frac{3\nu}{\sqrt{\Delta}}\left(\begin{array}{cc}
 v_{-,2} &  -1\\
- v_{+,2}& 1\end{array}  \right), 
\end{eqnarray}
 and consequently,
 \begin{eqnarray}
 e^{AN}=\frac{3\nu}{\sqrt{\Delta}}
 \left(\begin{array}{cc}
 e^{\lambda_+N}v_{-,2}-e^{\lambda_-N}v_{+,2} & e^{\lambda_-N}-e^{\lambda_+N}   \\
 & \\
 v_{+,2} v_{-,2}  (e^{\lambda_+N}-e^{\lambda_-N} )& e^{\lambda_-N}v_{-,2}-e^{\lambda_+N}v_{+,2}  \end{array}  \right). 
 \end{eqnarray}
This could be written in terms of the  discriminant as follows
  
\begin{eqnarray}
  e^{AN}=\frac{e^{-\frac{3}{2}(2+w_x-\nu+\mu)N}}{\sqrt{\Delta}}\left[
 \left(\begin{array}{cc}
 \sqrt{\Delta}\cosh\left(\frac{\sqrt{\Delta}N}{2}\right)& -6\nu\sinh\left(\frac{\sqrt{\Delta}N}{2} \right)  \\
 & \\
 6\mu \sinh\left(\frac{\sqrt{\Delta}N}{2} \right)  &       -\sqrt{\Delta}\cosh\left(\frac{\sqrt{\Delta}N}{2}\right)   
 \end{array}  \right)
 +3(w_x-\nu-\mu)\sinh\left(\frac{\sqrt{\Delta}N}{2} \right) {\rm Id}
 \right],\end{eqnarray}
where $\mbox{Id}$ denotes the identity matrix.

Finally, about the initial conditions it is useful to introduce the variables $\bar{\rho}_i=\frac{\rho_i}{3H_0^2M_{pl}^2}$ with $i=r, b, c, x$. Then, in the flat case $K=0$, the initial conditions are $\Omega_{i,0}$. We could choose the central value of $\Omega_{m,0}=0.31$ for the total matter (baryonic and dark) sector of the universe. Using the observational estimation of $\Omega_{b,0}h^2=0.0221$ and $\Omega_{c,0}h^2=0.1206$ we see that the percent of baryonic matter is approximately the $15.5\%$ of the total matter, so $\bar{\rho}_{b,0}=\Omega_{b,0}=0.048$ and $\bar{\rho}_{c,0}=\Omega_{c,0}=0.262$. Since, at the present time,  the energy density of radiation is negligible compared to other energy densities, one can approximately take
 $\bar{\rho}_{x,0}=\Omega_{x,0}=1-\Omega_{b,0}-\Omega_{c,0}\cong 0.69$.

To obtain the evolution of the energy density of radiation we 
 take, for example, the red-shift at the matter radiation-equality equal to
 its central value  $z_{eq}=3411$ for the PlanckTT+lowE likelihood \cite{Ade:2015xua}, for which $N_{eq}=-8.135$, and thus, from the matter-radiation equality,
 \begin{eqnarray}\label{initialconditions}
 \bar{\rho}_{r,eq}\equiv \bar{\rho}_{r}(N_{eq})=\bar{\rho}_{b}(N_{eq})+\bar{\rho}_{c}(N_{eq})= \bar{\rho}_{b,0}e^{-3N_{eq}}+\bar{\rho}_{c}(N_{eq}),  
 \end{eqnarray}
 once the parameters $\mu$, $\nu$ and $w_x$, are fixed, one obtains $\bar{\rho}_{r}(N_{eq})=\bar{\rho}_{r,eq} e^{-4(N-N_{eq})}$. And
 when the initial conditions are obtained one can easily calculate $\bar{\rho}_i(N)$ and also the effective Equation of State (EoS) parameter defined by 
 $w_{\rm eff}(N)= \frac{\rho_{tot}}{p_{tot}}=\frac{w_x\bar{\rho}_x(N)+\frac{1}{3}\bar{\rho}_r(N)}{\bar{\rho}_{tot}(N)}$.

\subsection{Special cases}
\label{sec-simple-cases}
 
In this section we consider the special cases emerging from the original interaction (\ref{modelA}). For example, one of the simplest cases that one may consider the case with $\nu=0$ in   (\ref{modelA}) which returns, $Q = 3 H \mu \rho_c$. Similarly, one could equally consider another case with $\mu=0$, equivalently, the interaction rate becomes $Q = 3 H \nu \rho_x$. The equality $\mu = \nu$ is also interesting. 
The solutions for all the cases are trivial, however, there is something that needs to be clarified in this article for future works. In what follows, we describe the solutions for each model as well as the bounds on the coupling parameters for which one obtains viable cosmological solutions.

 \begin{enumerate}
 \item For the interaction rate, $Q = 3 H \mu \rho_c$,  the condition (\ref{mu}) requires  $0.72w_x\leq \mu\leq 0$, and the solution, which is non-singular and positive, 
  is given by
 \begin{eqnarray}
 \bar{\rho}_c(N)=\Omega_{c,0}e^{-3(1+\mu) N}, \quad \mbox{and} \quad
 %\nonumber\\
 \bar{\rho}_x(N)=\Omega_{x,0}e^{-3(1+w_x)N}
 %+\nonumber \\ 
 -\frac{\mu\Omega_{c,0}}{\mu-w_x}\left( e^{-3(1+\mu) N}-e^{-3(1+w_x)N}\right). \end{eqnarray}
 
Finally, with this quantities one easily has the total pressure and energy density, so we have completely determined the effective EoS parameter $w_{eff}(N)$. It is easy to see that $w_{eff}(N)\rightarrow w_x$ when $N\rightarrow \infty$. To understand the evolution of various  components in terms of their energy densities in
Fig. \ref{fig:Omega1}, we display them for a specific choice of the coupling parameter, $\mu$ constrained in the region $0.72w_x\leq \mu\leq 0$ and for a particular choice of the EoS of DE, $w_x$. As one can see, all the energy densities remain positive for this choice of the coupling parameter. In Fig. \ref{fig:effw1}, we also show the evolution of the effective EoS, $w_{\rm eff}$ for the same value of $\mu$ and $w_x$ used to draw Fig. \ref{fig:Omega1}. To be precise, in the left panel of Fig. \ref{fig:effw1}, we show the evolution of $w_{\rm eff}$ for  a wide region where $N \in [-10, 60]$ and in the right panel of Fig. \ref{fig:effw1} we show the evolution of $w_{\rm eff}$ from the epoch of matter-radiation equality to present time. One can clearly visualize from Fig. \ref{fig:effw1}, that $w_{\rm eff}$ crosses from positive (in the early phase of the universe) to negative values (at present time) and then asymptotically converges to $w_x = -0.95$. 
 
\begin{figure}[H]
\begin{center}
\includegraphics[scale=0.45]{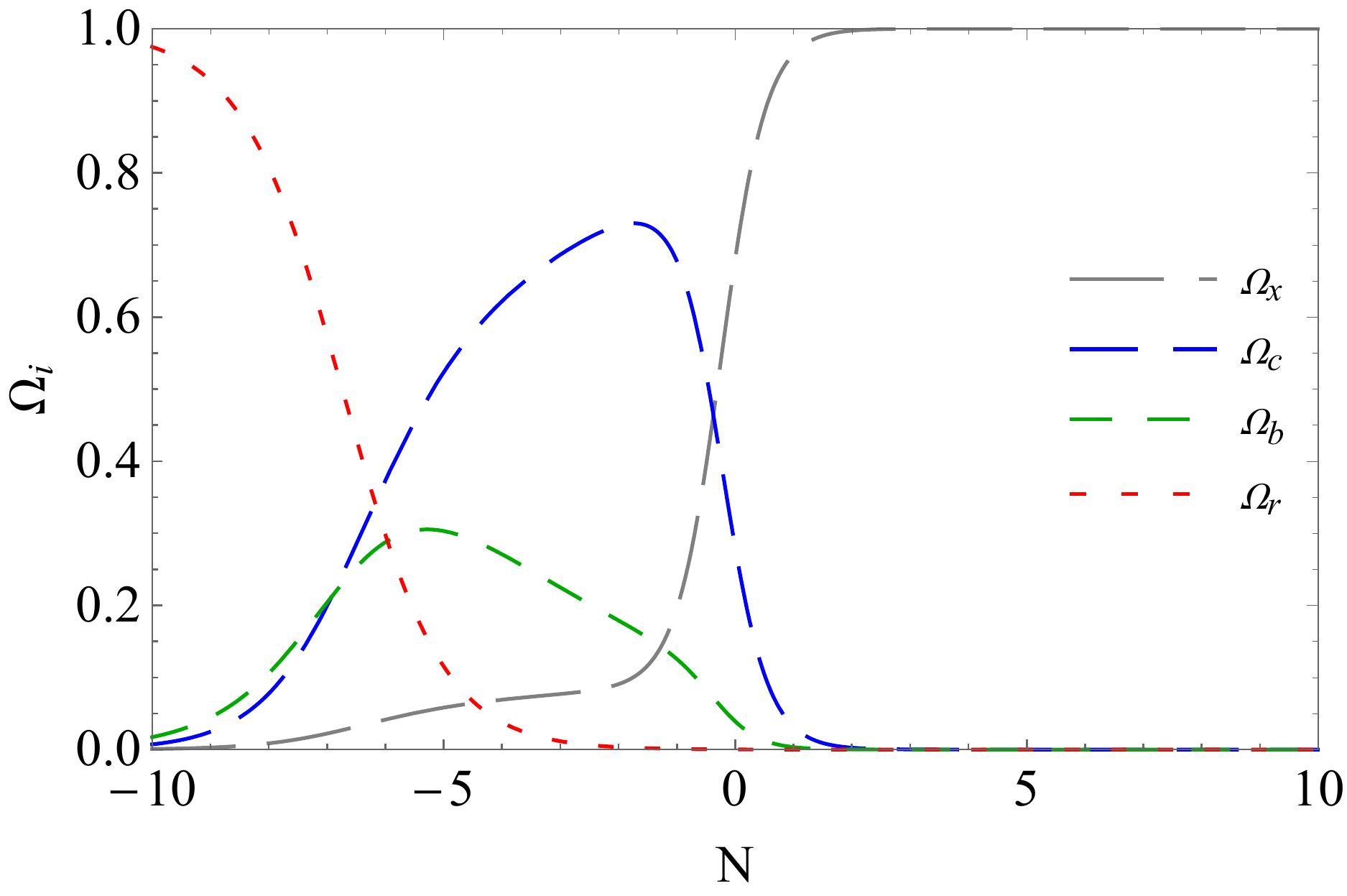}
\end{center}
\caption{The evolution of the density parameters namely, $\Omega_r$ (red curve), $\Omega_b$ (green curve), $\Omega_c$ (blue curve) and $\Omega_x$ (grey curve)  for the interaction rate (\ref{modelA}) with $\nu=0$, $\mu=0.1w_x$, $w_x=-0.95$ has been shown in this picture.  }
\label{fig:Omega1}
\end{figure}
\begin{figure}[H]
\begin{center}
\includegraphics[scale=0.35]{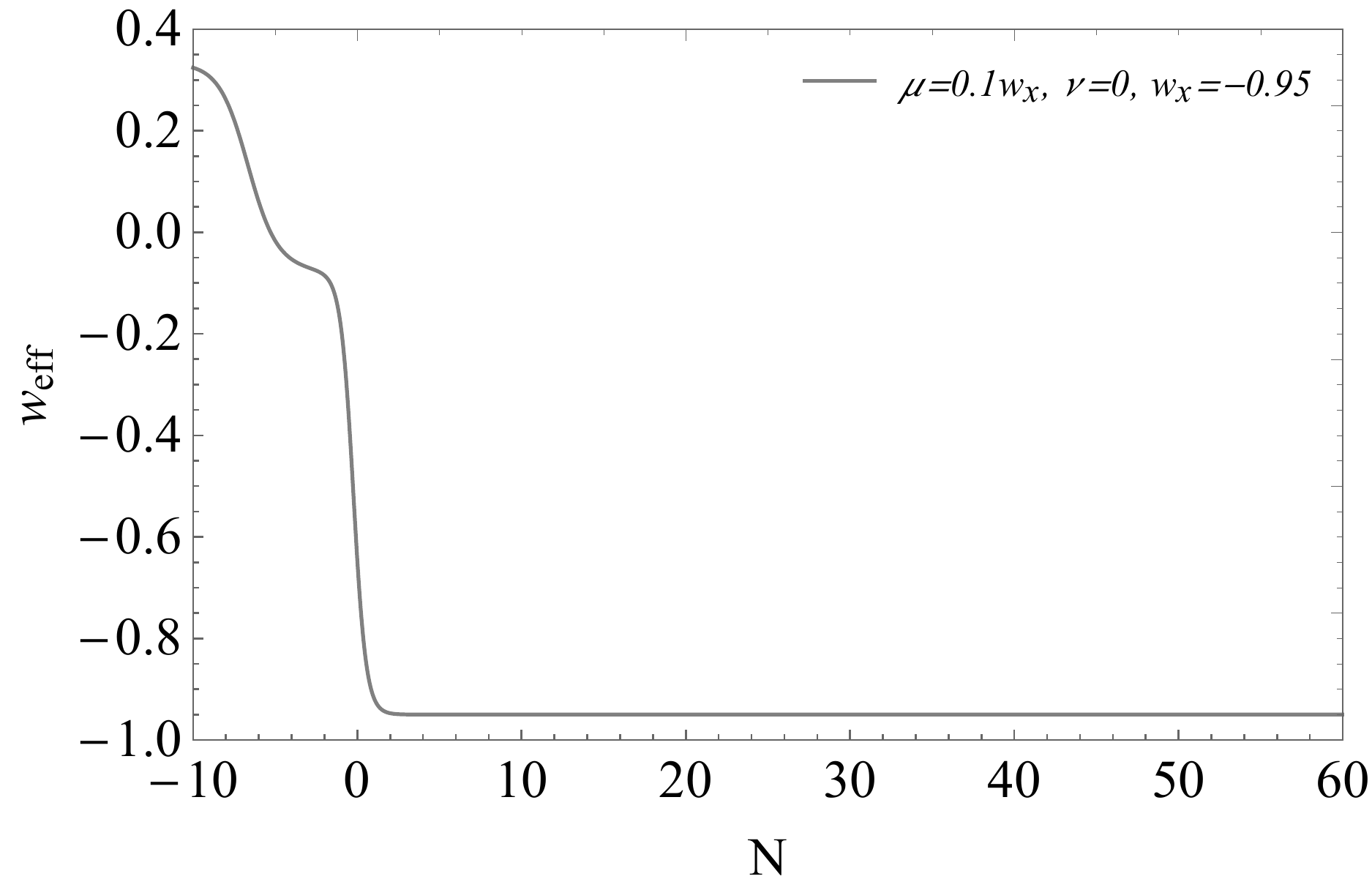}
\includegraphics[scale=0.35]{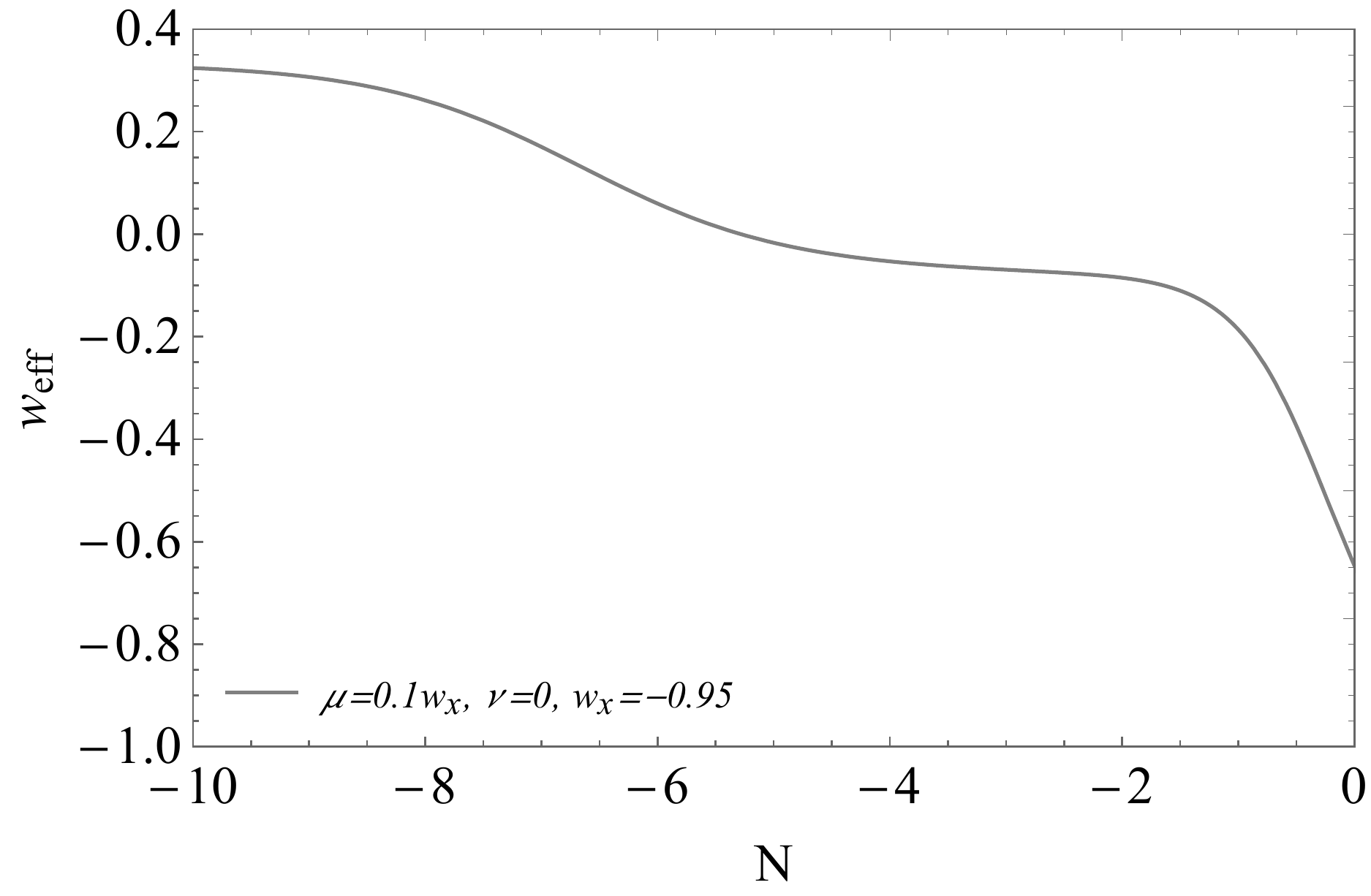}
\end{center}
\caption{We show the evolution of the effective EoS of DE for the interaction scenario with $\nu=0$, $\mu=0.1w_x$, $w_x=-0.95$.  In the left panel we see the asymptotic evolution of $w_{\rm eff}$ for the region $N \in [-10, 60]$, and in the right panel we only show its evolution from the matter-radiation equality to the present time, that means for $N \in [-10, 0]$. One can notice that the effective equation of state parameter converges to $w_x=-0.95$. }
\label{fig:effw1}
\end{figure}

 \item   For the second case with $\mu=0$, equivalently for the interaction rate $Q  = 3 H \nu \rho_x$, working in the same way as above, one  obtains $\lambda_+=-3(1+w_x-\nu)$ and $\lambda_-=-3$, and thus,
 \begin{eqnarray}
 v_+=\left(1,\frac{w_x-\nu}{\nu}\right)\qquad v_-=(1,0).
 \end{eqnarray}
 
 In that case the condition $v_{\pm,1}v_{\pm, 2}\geq 0$  and $\frac{\rho_{x,0}}{\rho_{c,0}}\leq \frac{w_x-\nu}{\nu}$
requires  $0.27w_x\leq \nu \leq 0$, and  
  the solution is given by
 \begin{eqnarray} \label{wx}
 \bar{\rho}_x(N)=\Omega_{x,0}e^{-3(1+w_x-\nu)N}\quad \mbox{and} \quad \bar{\rho}_c(N)=\Omega_{c,0}e^{-3 N}
 +\frac{\nu\Omega_{x,0}}{w_x-\nu}\left(e^{-3(1+w_x-\nu)N}-e^{-3N}   \right).
 \end{eqnarray}
 
Finally,  a simple calculation shows that
 \begin{eqnarray}
 \lim_{N\rightarrow \infty}w_{eff}(N)\equiv w_{eff,\infty}=w_x-\nu\geq w_x,
 \end{eqnarray}
 but this does not mean that the universe could decelerate once again.
 Effectively, at the present time we have $w_{eff,0}\cong w_x\Omega_{x,0}$ because the radiation is negligible. Then, since  nowadays our universe is
 accelerating, hence, $w_{eff,0}$ has to be less than $-1/3$ and taking a typical value of the density parameter for DE as $\Omega_{x,0}\cong 0.69$, this  means that $-1\leq w_x< 0.483$, and thus,
 \begin{eqnarray}
 w_{eff,\infty}<(1-0.27)w_x=0.73 w_x<-0.73\times 0.483\cong -0.35 <-1/3,
 \end{eqnarray}
 meaning that the universe accelerates at late times. Similarly, 
for this special case too, we have calculated the density parameters of different cosmic fluids as well as the effective EoS of the total fluid.  In 
Fig. \ref{fig:Omega2}, we show the density parameters for a specific choice of the coupling parameter, $\nu$ satisfying the constraint $0.72w_x\leq \nu \leq 0$ and for a particular choice of the EoS of DE, $w_x = -0.95$. Additionally, in Fig. \ref{fig:effw2}, we depict the evolution of the effective EoS, $w_{\rm eff}$ for the same value of $\nu$ and $w_x$ used to draw Fig. \ref{fig:Omega2}. In the left panel of Fig. \ref{fig:effw2}, we describe the evolution of $w_{\rm eff}$ for $N \in [-10, 60]$ while in the right panel of Fig. \ref{fig:effw2} we show the evolution of $w_{\rm eff}$ from the epoch of matter-radiation equality to present time. A clear transition of $w_{\rm eff}$ from its positive values to negative values are found from Fig. \ref{fig:effw2}. 

 \begin{figure}[H]
\begin{center}
\includegraphics[scale=0.45]{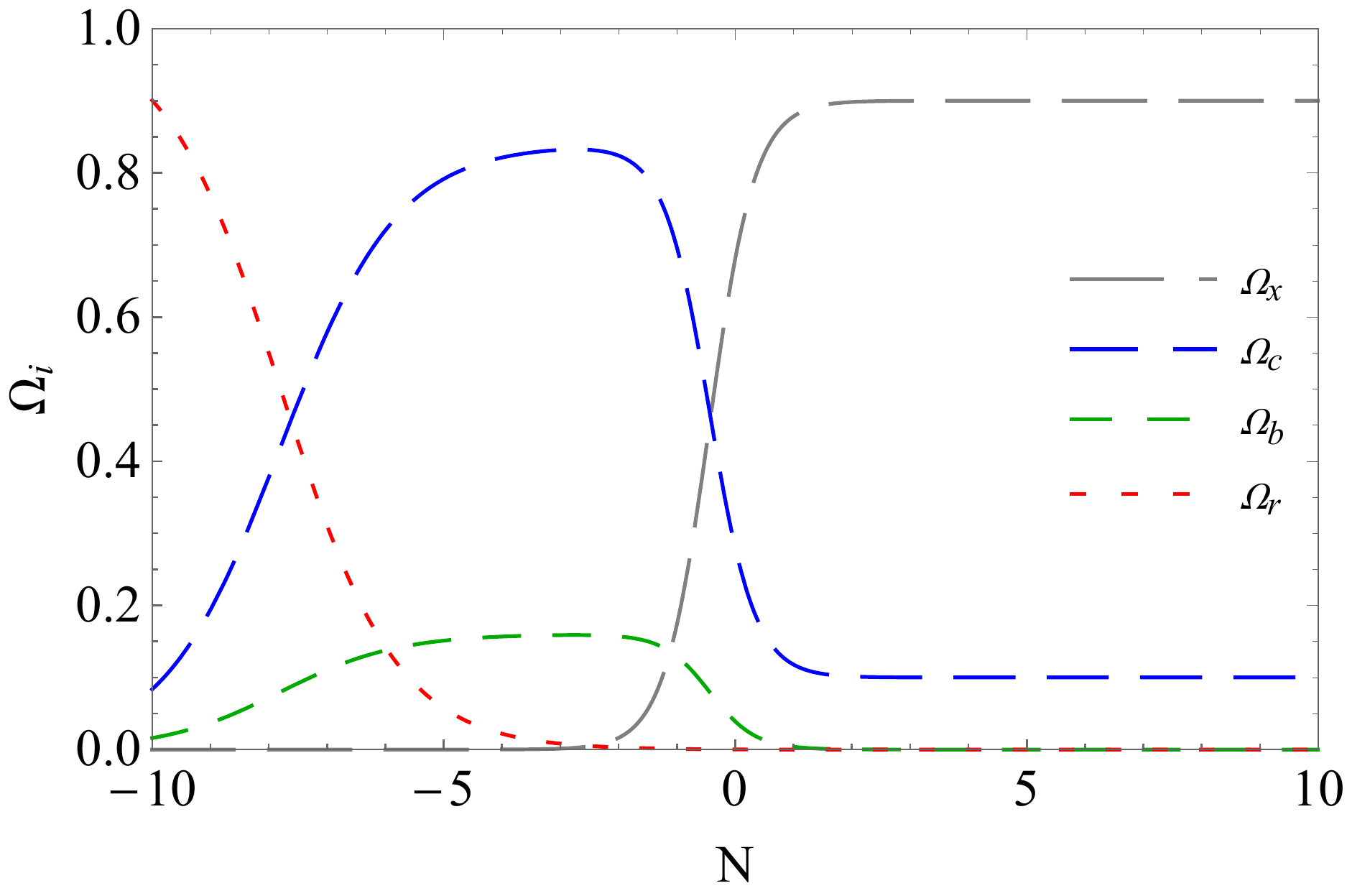}
\end{center}
\caption{The evolution of the density parameters namely, $\Omega_r$ (red curve), $\Omega_b$ (green surve), $\Omega_c$ (blue curve) and $\Omega_x$ (grey curve) for the interaction rate (\ref{modelA}) with  $\mu=0$, $\nu=0.1w_x$, and $w_x=-0.95$ has been displayed.  }
\label{fig:Omega2}
\end{figure}
\begin{figure}[H]
\begin{center}
\includegraphics[scale=0.35]{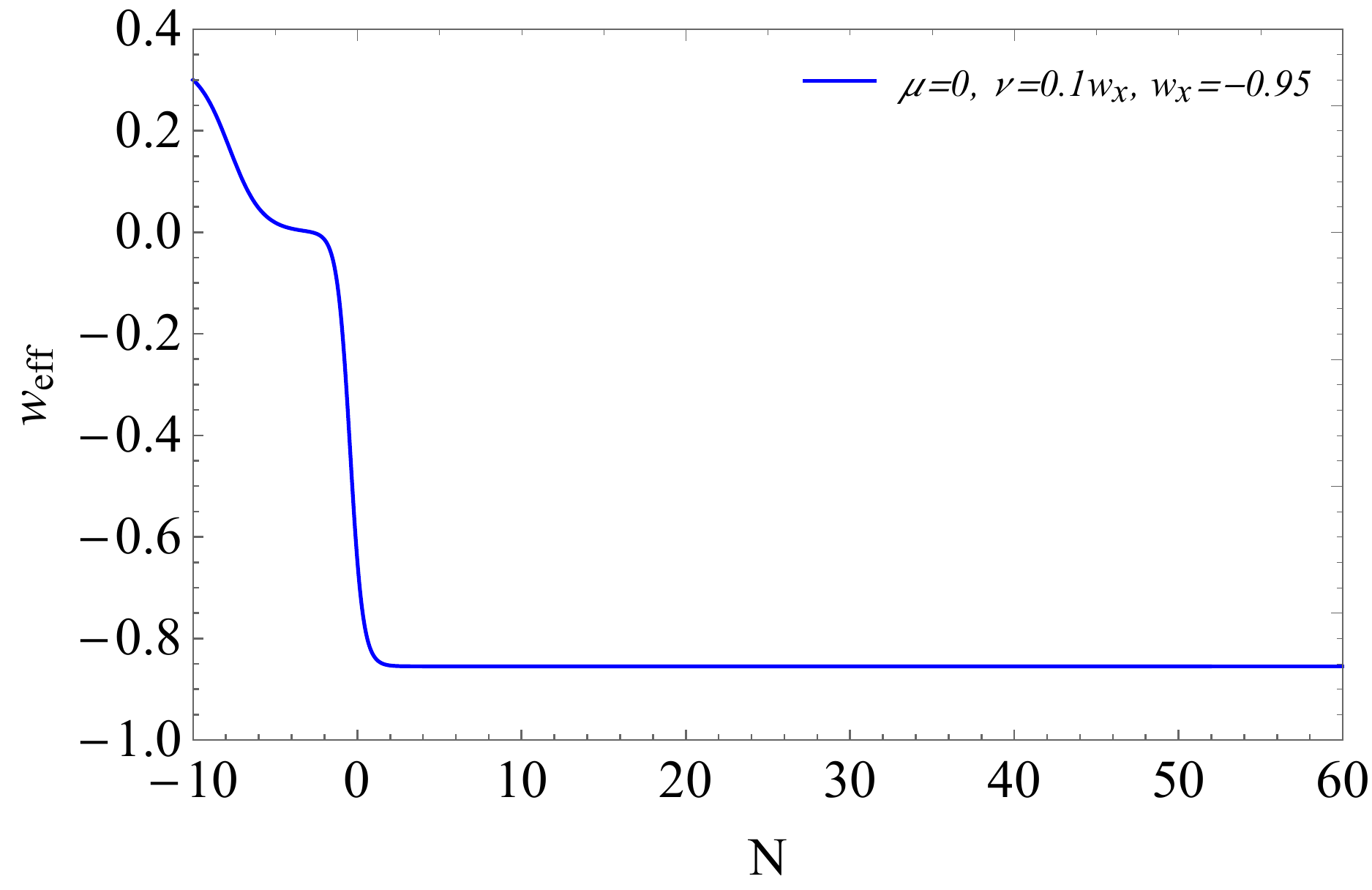}
\includegraphics[scale=0.35]{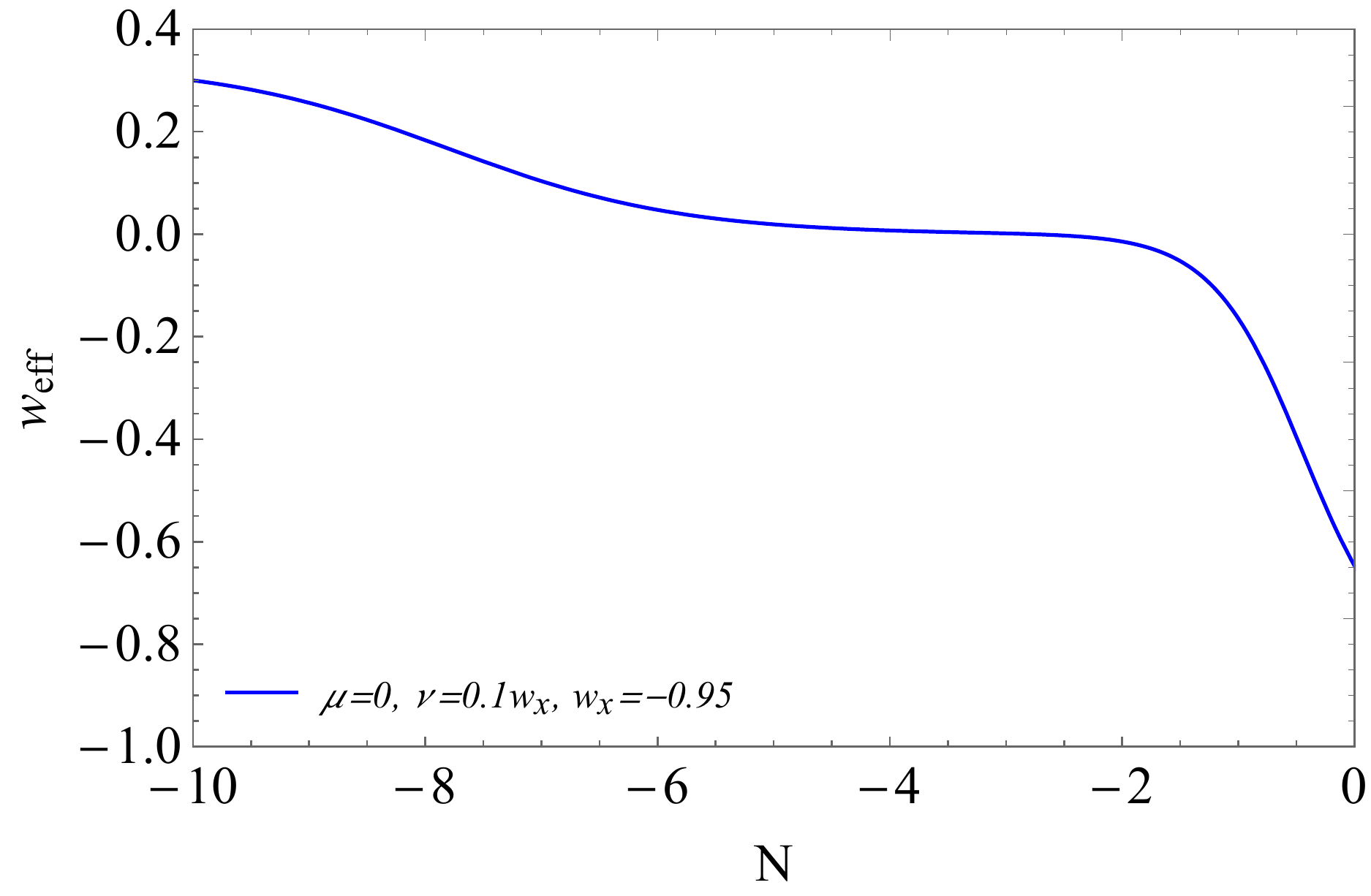}
\end{center}
\caption{We display the evolution of the effective EoS of DE for the interaction rate (\ref{modelA}) with $\mu=0$, $\nu=0.1w_x$, $w_x=-0.95$.  In the left panel we see the asymptotic evolution of $w_{\rm eff}$ for the region $N \in [-10, 60]$, and in the right panel we only show its evolution from the matter-radiation equality to the present time, that means for $N \in [-10, 0]$. One can notice that the effective equation of state parameter converges to $w_x-\nu=-0.855$ in an asymptotic manner. }
\label{fig:effw2}
\end{figure}

{\bf Remark 1:} {\it  We would like remark that to obtain a viable cosmological scenario driven by the linear interaction models prescribed above, one needs to impose the condition $0.27w_x\leq \nu \;(\mbox{or}\; \mu) \leq 0$. However, sometimes this condition is somehow overlooked and due to arbitrary choice of the coupling parameters, the energy densities of the dark sector do not remain positive throughout the  evolution of the universe. Sometimes a negative sign is considered in the interaction rate which leads to additional confusions. Thus, we would like to clarify this point in our notation. For instance, let us select the interaction rate  $Q=-\xi H \rho_x$, which in our notation implies, $\xi = -3 \nu$. Now, since $0.27w_x\leq \nu \leq 0$, hence, we need to choose positive values of $\xi$. The energy densities become positive for times previous to the present time, but  as we can see from (\ref{wx}) the energy density of the dark matter becomes negative at late times. Alternatively, if we select  $Q=\alpha H\rho_c$ which in our notation $\mu=\alpha/3$. Again we need to choose $\alpha <0$ because for positive values of $\alpha$, the energy density of the dark energy $\rho_x$ becomes negative at early times, meaning that no viable cosmic scenario for these parameters.  }

\item Another theoretically interesting case might be the one when the coupling parameters are equal, that means $\mu=\nu$.  In this situation the trace is given by $\mbox{Tr}\; A=-3(2+w_x)$, and it is negative because we are considering non-phantom fluids. The determinant is given by $\mbox{Det}\;A=9(1+w_x+\mu w_x)$, and it is positive when $\mu<-1-\frac{1}{w_x}$. For the discriminant one has
 $\Delta=9w_x(w_x-4\mu)$, meaning that it is positive for $\mu>w_x/4$. So, for the moment we have $w_x/4<\mu<-(w_x+1)/w_x$.
 Now we deal with the condition (\ref{eigenvector}). Since
 \begin{eqnarray}
 \lambda_{\pm}=-\frac{3(2+w_x)}{2}\pm\frac{3}{2}\sqrt{(2\mu-w_x)^2-4\mu^2},
 \end{eqnarray}
 the constraint (\ref{eigenvector}) becomes
 \begin{eqnarray}
 \frac{1}{2\mu}\left(2\mu-w_x\pm \sqrt{(2\mu-w_x)^2-4\mu^2}\right)\leq 0, 
 \end{eqnarray}
 which is satisfied only for $w_x/2\leq \mu\leq 0$. Then, together with $w_x/4<\mu<-(w_x+1)/w_x$, we will have  
 \begin{eqnarray}
 w_x/4< \mu\leq 0.
 \end{eqnarray}

 Finally, the condition (\ref{energydensities0}) becomes
 \begin{eqnarray}
-\frac{1}{2\mu}\left(2\mu-w_x- \sqrt{(2\mu-w_x)^2-4\mu^2}\right)\leq 
\frac{\rho_{x,0}}{\rho_{c,0}}\leq
-\frac{1}{2\mu}\left(2\mu-w_x+ \sqrt{(2\mu-w_x)^2-4\mu^2}\right).\end{eqnarray}
  \end{enumerate}
 
\begin{figure}[H]
\begin{center}
\includegraphics[scale=0.45]{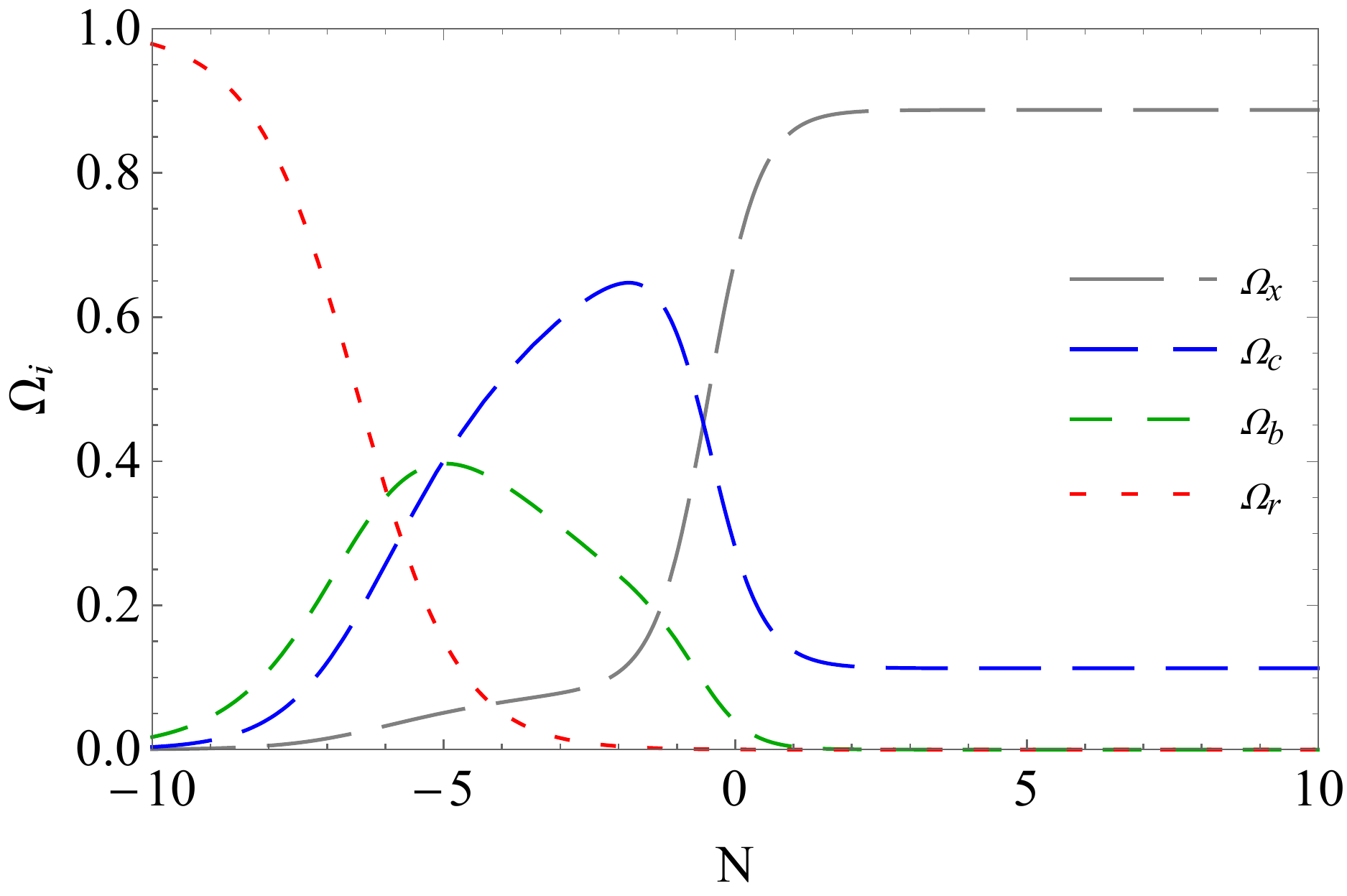}
\end{center}
\caption{The evolution of the density parameters, namely, $\Omega_r$ (red curve), $\Omega_b$ (green curve), $\Omega_c$ (blue curve) and $\Omega_x$ (grey curve) for the  interaction rate (\ref{modelA}) with $\mu=\nu=0.1w_x$ and  $w_x=-0.95$ has been shown.  }
\label{fig:Omega3}
\end{figure}

\begin{figure}[H]
\begin{center}
\includegraphics[scale=0.35]{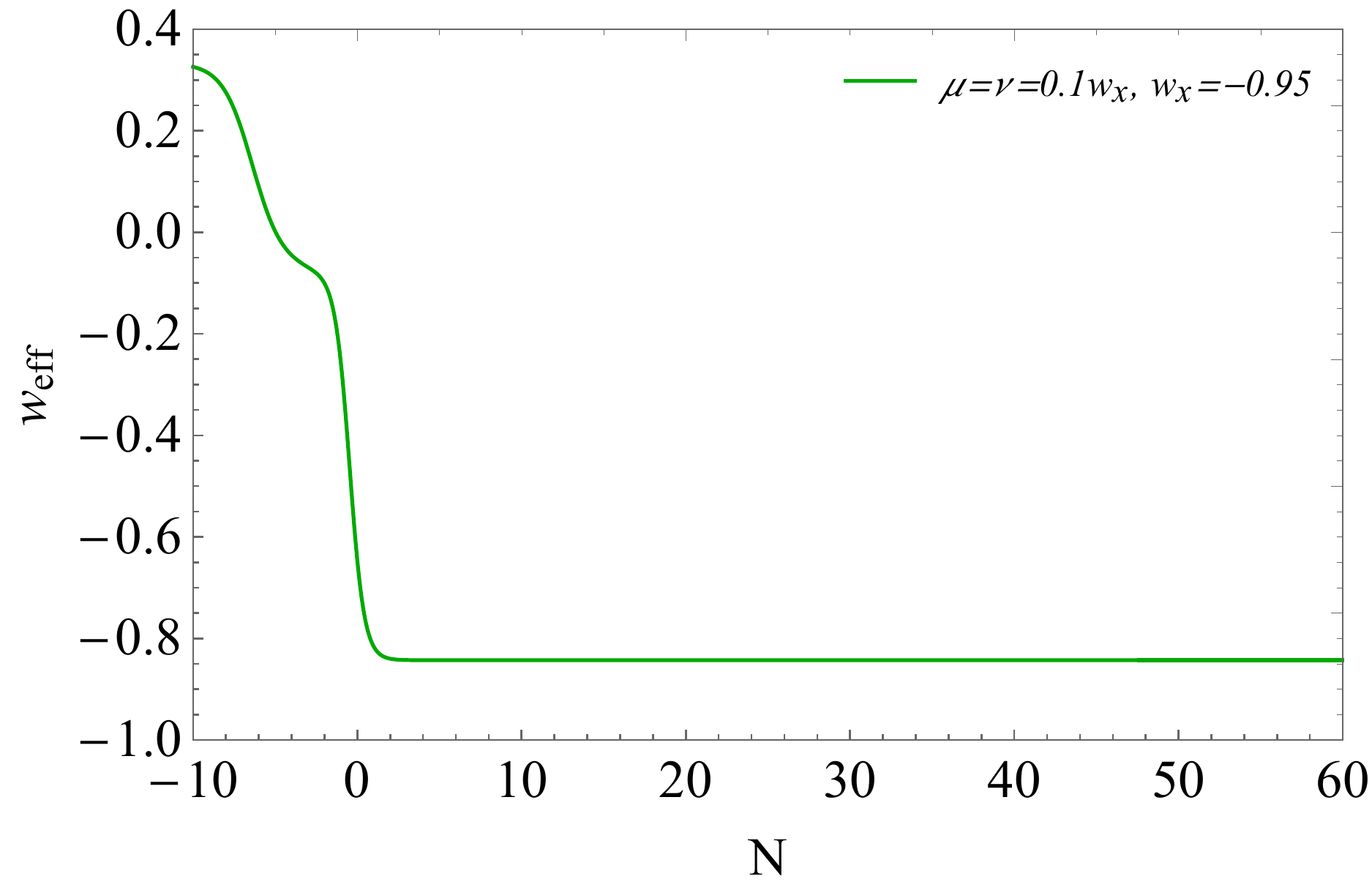}
\includegraphics[scale=0.35]{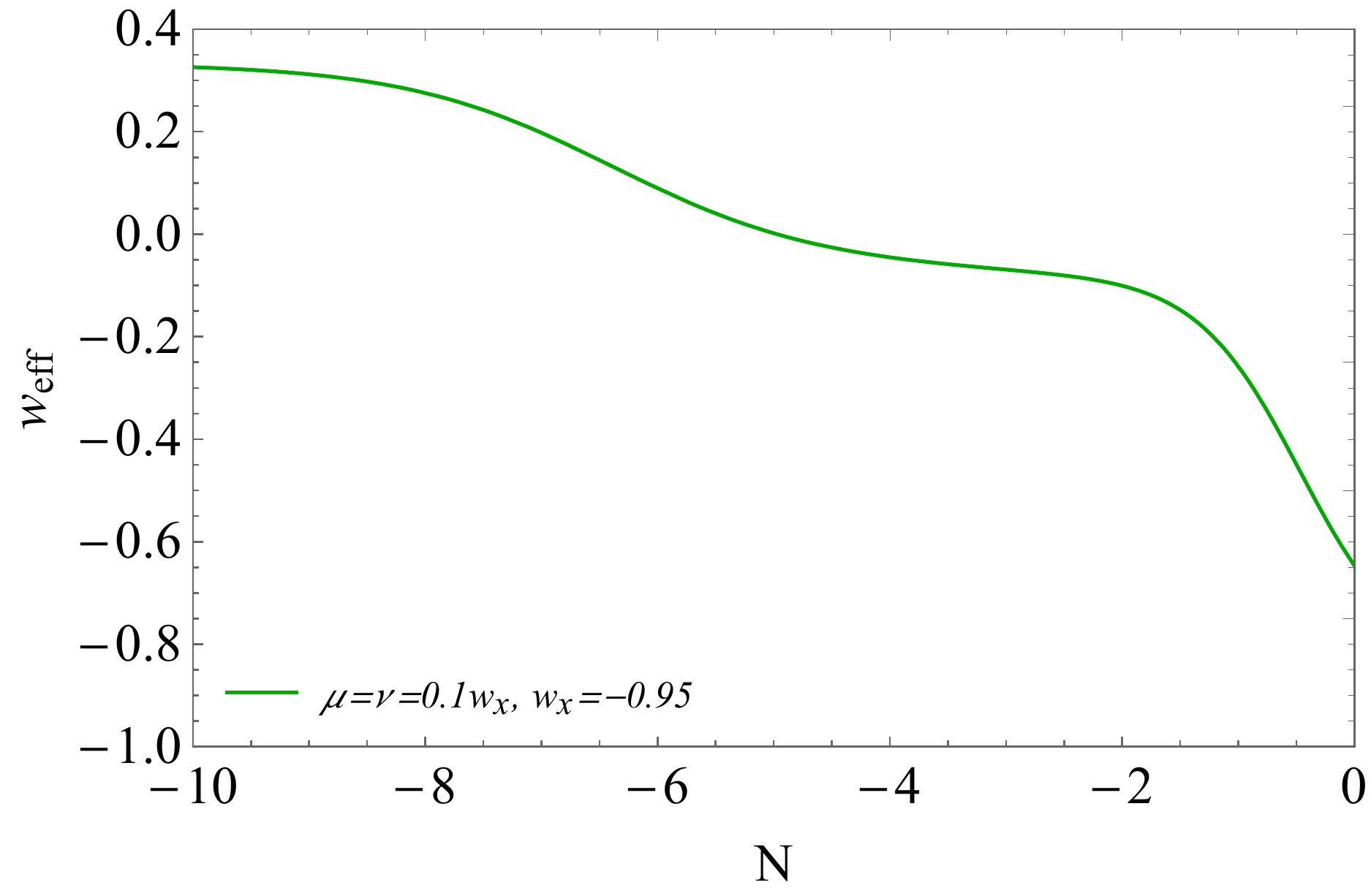}
\end{center}
\caption{We display the evolution of the effective EoS of DE for the interaction rate (\ref{modelA}) with $\mu=\nu=0.1w_x$, $w_x=-0.95$.  In the left panel we show the asymptotic evolution of $w_{\rm eff}$ for the region $N \in [-10, 60]$, and in the right panel we only show its evolution from the matter-radiation equality to the 
present time, that means for $N \in [-10, 0]$.  }
\label{fig:effw3}
\end{figure}

\begin{figure}[H]
\begin{center}
\includegraphics[scale=0.45]{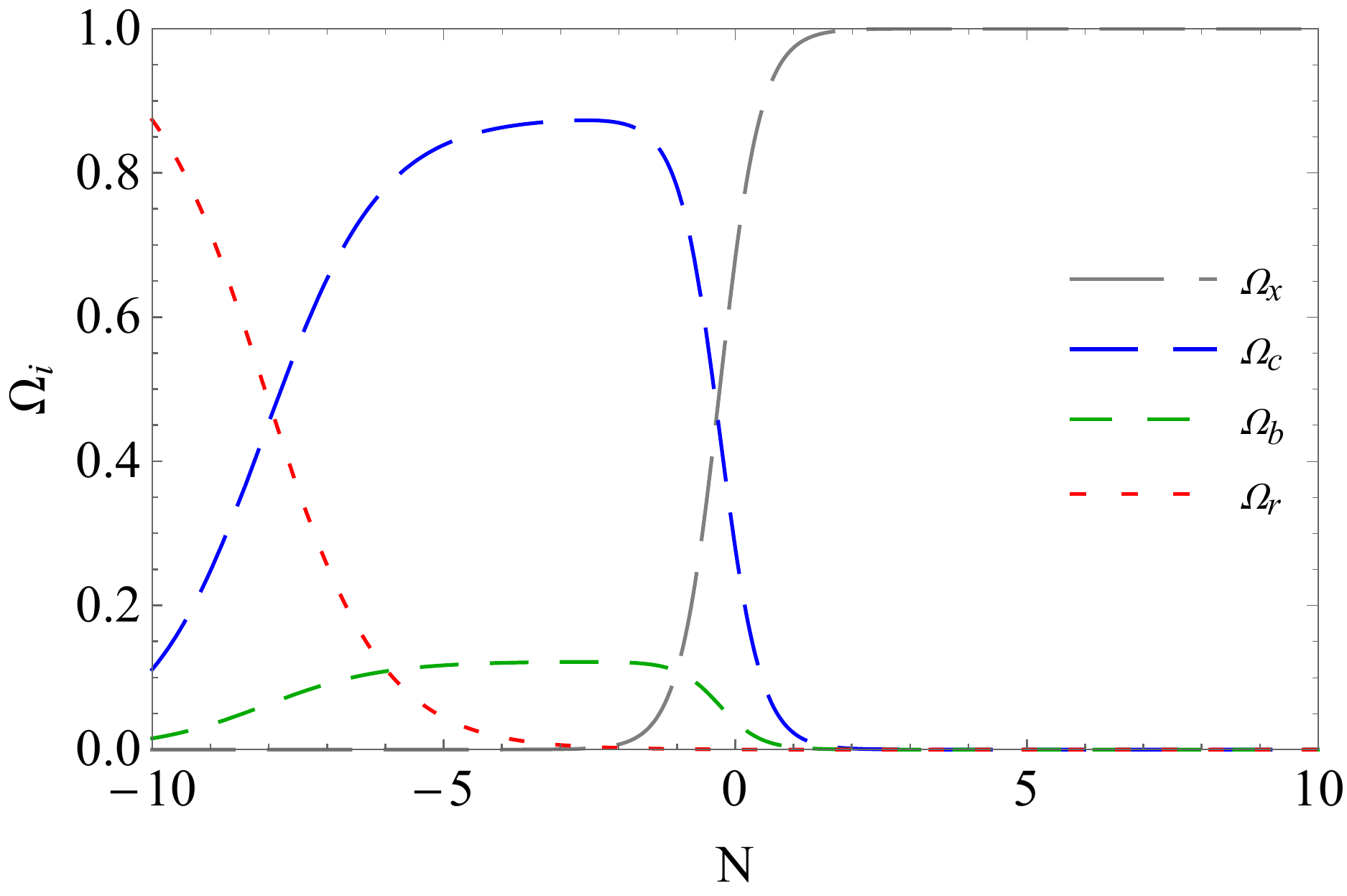}
\end{center}
\caption{We show the evolution of the density parameters, namely, $\Omega_r$ (red curve), $\Omega_b$ (green curve), $\Omega_c$ (blue curve) and $\Omega_x$ (grey curve) for the non-interacting case $\mu=\nu=0$, with $w_x=-0.95$. }
\label{fig:Omega4}
\end{figure}

\begin{figure}[H]
\begin{center}
\includegraphics[scale=0.35]{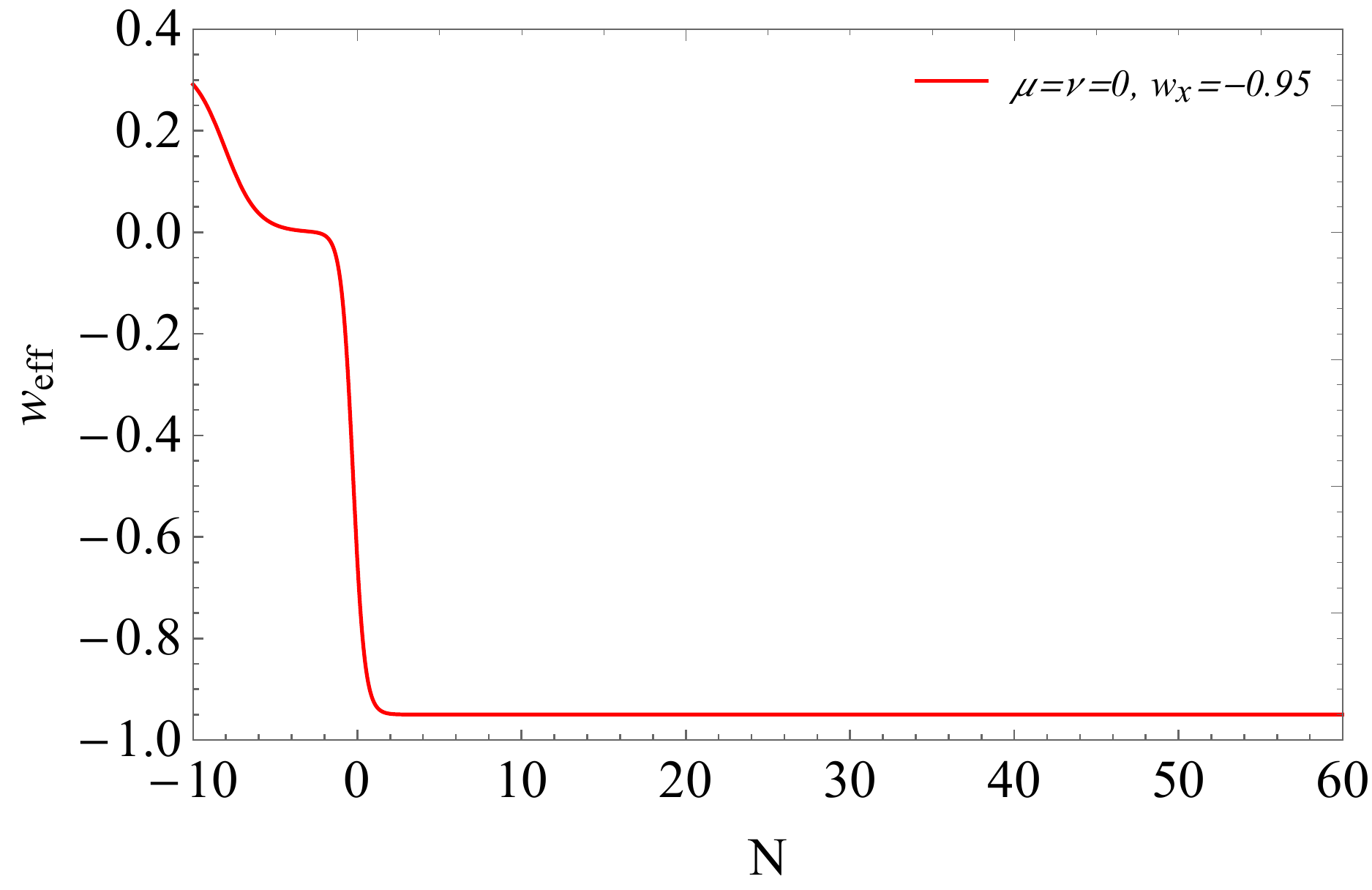}
\includegraphics[scale=0.35]{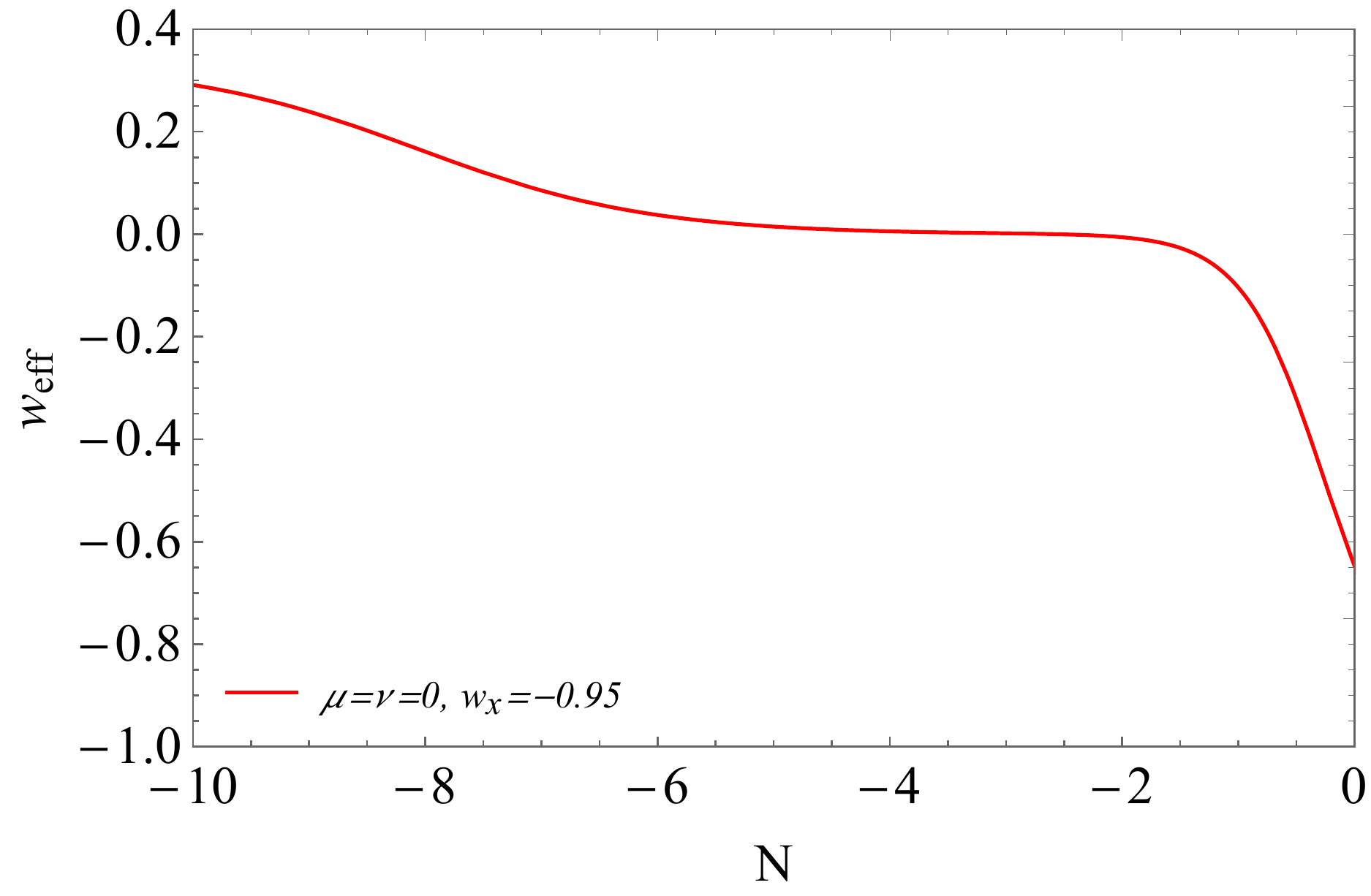}
\end{center}
\caption{We display the evolution of the effective EoS of DE for the non-interaction scenario that means with $\mu= \nu =0$, and $w_x=-0.95$. In the left panel we show the asymptotic evolution of $w_{\rm eff}$ for the region $N \in [-10, 60]$, and in the right panel we only show its evolution from the matter-radiation equality to the 
present time, that means for $N \in [-10, 0]$. }
\label{fig:effw4}
\end{figure}

 Now since $\frac{\rho_{x,0}}{\rho_{c,0}}= \frac{\Omega_{x,0}}{\Omega_{c,0}} \cong 2.63$ we  obtain the following bound
 \begin{eqnarray}
 0.18w_x\leq \mu\leq 0 \Longrightarrow -0.18\leq \mu\leq 0,
 \end{eqnarray}
 because as already mentioned, we are dealing with non-phantom fields, i.e., $ -1 \leq w_x < -1/3$. For this interaction model we have similarly shown the modified density parameters  in Fig. \ref{fig:Omega3} and the effective equation of state parameter in Fig. \ref{fig:effw3} taking the choices $\mu = nu = 0.1 w_x$ with $w_x  =-0.95$.  Finally, we have considered a non-interacting scenario of the universe, equivalently, $\mu = \nu = 0$ and shown the density parameters as well as the effective equation of state in Figs. \ref{fig:Omega4} and \ref{fig:effw4}, respectively, taking a specific value of $w_x  =-0.95$. The Figs. \ref{fig:Omega4} and \ref{fig:effw4} are motivated to present a comparison between various interacting scenarios with the non-interacting one.


It is well known that in the non-interacting case the perturbations grow in a matter-dominated regime, i.e., when $w_{\rm eff}=0$ and the physical scales re-enter the Hubble horizon $k^2\gg a^2H^2$ \cite{Mukhanov}  (see also  \cite{doran} where the authors, using a matrix approach, show that, in the non interacting case, the dominant perturbations become constant in the super-horizon scales during the radiation era, and for an interacting case, see  \cite{li}, where the authors, following the same analysis as in \cite{doran}, arrive at the same conclusion). So, a matter-dominated state is necessary after the radiation domination and before the present time. Although this is a topic that deserves future investigations, we hope that the same will happen in the interacting case for values of the parameters $\mu$ and $\nu$ close to zero. For this reason,  comparing the different effective equation of state parameters studied here, we see that the best case is when $\mu=0$ and $\nu=0.1w_x$ with $w_x=-0.95$.

In fact, studies  in \cite{Valiviita:2008iv} agree that interacting models may lead to early time instabilities where the instabilities of the perturbations are only considered in super-horizon scales and extrapolated (see for instance equations (63)--(75) of  \cite{Valiviita:2008iv}) to sub-horizon scales. However, why one can extrapolate from super-horizon to sub-horizon scales, is not clear, and hence, the instability problem in this context does not seem to be conclusive and this deserves further investigations.  Moreover, when obtaining perturbations many orders greater than the background, it has not been realized that the linear approximation only holds for perturbations less or the same order of the background.  
More interesting are the conclusions of \cite{Olivares}, where the authors state the difficulty of the study of sub-horizon perturbations because it is impossible, in this regime, to find analytic solutions, and the only  conclusion is  that the perturbations grow slower than in the non-interacting case.  The understanding of early time instabilities appearing in the interacting DE-DM theories is therefore a key topic that deserves serious attention. Such instabilities are highly model dependent since it can be avoided with the proper choice of the interaction function \cite{li}. So, one may argue that the instabilities appearing in such theories point toward the insufficiencies of the phenomenological parameterizations of the interaction functions.  Although the linear interaction model is the most simplest and convenient choice to proceed with, however, one may consider suitable parameterizations of the interaction functions beyond the linear parameterizations with an aim to investigate the evolution of the interacting universe at the level of perturbations.

%&&&&&&&&&&&&&&&&

\section{Future singularities}
\label{sec-singularities}

In this section we demonstrate that the interaction scenarios may lead to finite time singularities. In particular, we find that our present 
linear interacting model may also lead to finite time future singularities. Effectively, if one takes $\mbox{Det}\; A =9(\mu+1)(1+w_x-\nu)+9\mu\nu<0$, the origin becomes a saddle point, and thus, at very late times, the total energy density diverges, meaning that  the universe will enter in a phantom dominated phase finishing in a BIG RIP singularity. To show that, we choose for example, $\mu=0$ which implies $1+w_x-\nu<0$ and this restriction together with the condition 
 $\frac{\rho_{x,0}}{\rho_{c,0}}\leq \frac{w_x-\nu}{\nu}$ leads to the constraint
 $$\mbox{max}(0.27w_x,1+w_x)<\nu<0,
 $$
 which implies that $w_x<-1$, that means, effectively $\rho_x$ must be a phantom fluid. Thus, at late times, the solution (\ref{wx}) becomes
 $$
 \bar{\rho}_x(N)\sim \Omega_{x,0}e^{-3(1+w_x-\nu)N}\bar{\rho}_x(N)\rightarrow \infty,\quad \mbox{and} \quad \bar{\rho}_c(N)\sim 
 \frac{\nu}{w_x-\nu}\bar{\rho}_x(N) \bar{\rho}_x(N)\rightarrow \infty, $$
and  $w_{eff,\infty}=w_x-\nu<-1$, which means that, at late times, the universe enters into a phase dominated by the phantom fluid. 
 
 Finally, at very late times, the Raychaudhuri equation for $K =0$ universe becomes
 $$
 \dot{H}=-\frac{1}{2M_{pl}^2}(1+w_{eff,\infty})\rho_{tot}=-\frac{3}{2}(1+w_{eff,\infty})H^2, $$
 whose solution is given by
 $$
 \frac{1}{H_0}-\frac{1}{H}=-\frac{3}{2}(1+w_{eff,\infty})(t-t_0) \Longleftrightarrow  H(t)=\frac{2}{3(1+w_{eff,\infty})}\frac{1}{t-t_s},
 $$
 with $t_s=t_0-\frac{2}{3H_0(1+w_{eff,\infty})}>t_0$. This means that the universe has a finite time cosmic singularity in the future, where the total energy density and pressure diverge, i.e., the model has a BIG RIP singularity.

%&&&&&&&&&&&&&&&&&&&&&&&&&&&&&&&&&&&&&&&&&&&&

On the other hand, when $\mbox{Det}\; A>0$ and
$\mbox{Tr}\; A>0$, both the eigenvalues are positive, and the origin of coordinates is a repeller, which means that one also obtains a BIG RIP singularity at late times. In order to ensure that the energy densities of the fluids were always positive, one has to impose that origin was not a focus, because if so, at early times the orbits would oscillate around the origin leading to negative energies. To prevent this  behavior, one has to impose that the discriminant $\Delta = (\mbox{Tr}\; A)^2- 4 \mbox{Det}\; A$ was positive, i.e., the origin was a node. In addition, as we have already explained in Section \ref{sec-dyn-analysis}, for a node, to guarantee the positive values  of  the energy densities, both the orbits following the respective  eigenvectors of the matrix $A$, ($X_+(N)=e^{\lambda_+N}v_+$ and $X_-(N)=e^{\lambda_-N}v_{-}$ being once again  $\lambda_+$ and $\lambda_-$ the eigenvalues of the matrix $A$ and, $v_+=(v_{1,+},v_{2,+})$ and $v_-=(v_{1,-},v_{2,-})$ their corresponding eigenvectors) must belong to the first quadrant, that is, the condition $v_{1,\pm}v_{2,\pm}>0$ must be satisfied. 
As a consequence, all orbits with an initial  condition defined in this region must remain in this region of the first quadrant, which ensures that the energy densities were always positive. 

\

All these conditions lead to the following constraints that  the parameters $\mu$ and $\nu$ need to satisfy:
\begin{equation}\label{condition}
    \left\{\begin{array}{ccc}
      2+\mu-\nu+w_{x}   &< & 0  \\
    (1+\mu)(1+w_{x})-\nu     & >&0\\
    (w_{x}-\mu-\nu)^2-4\mu\nu &> & 0,
    \end{array}\right.
\end{equation}
where the first equation implies that $\mbox{Tr}\; A>0$, the second implies $\mbox{Det} A>0$ and the third ensures $\Delta>0$. 
Moreover, as we know, the eigenvalues of $A$ are given by $\lambda_{\pm}=
(\mbox{Tr}A\pm
\sqrt{\Delta})/2$ and the corresponding eigenvectors can be classified for $\nu \neq 0$ and $\nu = 0$ as follows: 

\begin{enumerate}
    \item For $\nu\not=0$, the eigenvectors are given by
    \begin{equation}
        v_{\pm}=\left(1, -\frac{\mu+1+\lambda_{\pm}/3}{\nu}\right).
    \end{equation}
    \item For $\nu=0$, the eigenvalues are $\lambda_+=-3(1+w_{x})$ and 
    $\lambda_-=-3(1+\mu)$, which implies that $w_{x}<-1$ (phantom fluid) and also $\mu<-1$. The eigenvectors are then given by
    \begin{equation}
        v_+=(0,1), \quad v_-=\left(1, \frac{\mu}{w_{x}-\mu} \right).
    \end{equation}
\end{enumerate}

Now, the condition $v_{1,\pm}v_{2,\pm}>0$ gives
\begin{equation}\label{a}
    \left\{\begin{array}{ccc}
   \frac{\mu+1+\lambda_{\pm}/3}{\nu}>0      &\mbox{for} &\nu\not=0 \\
   & &\\
   w_{x}\leq \mu<0   & \mbox{for}  &  \nu=0.
    \end{array}\right.
\end{equation}

Thus, in order that the initial condition was in the region defined by the orbits $X_+(N)$ and $X_-(N)$, and thus, the energy densities were always positive,  we get the following restrictions: 
\begin{equation}\label{b}
   \mbox{min}(v_{2,+}, v_{2,-})\leq  \frac{\rho_{x,0}}{\rho_{c,0}}\leq \mbox{max}(v_{2,+}, v_{2,-}),\quad \mbox{for}\; \nu \neq 0,
\end{equation}
 and 
\begin{equation}\label{nu0}
    \frac{\rho_{x,0}}{\rho_{c,0}}>
    \frac{\mu}{w_{x}-\mu}, \quad \mbox{for}\; \nu =0.
\end{equation}

Then, considering the simple case $\nu=0$, and taking into account that $\rho_{x,0}/\rho_{c,0}=
\Omega_{x,0}/\Omega_{c,0}$,
and using, once again, the following observational values at present time, namely, $\Omega_{c,0}\cong 0.262$ and
$\Omega_{x,0}\cong 0.69$, from (\ref{nu0}) we deduce that the parameter $\mu$ must satisfy 
the condition
\begin{equation}
    0.72w_{x}<\mu<0,
\end{equation}
and from (\ref{condition}) we conclude that, to have a future BIG RIP singularity,  the value of the parameter $\mu$ has to satisfy
\begin{equation}
0.72w_{x}<\mu<-1, \quad \mbox{ with }\quad w_{x}<-1.38.
\end{equation}

With the analysis presented above, it is clear that  the
linear interaction models may encounter with finite time future singularities (here the BIG RIP singularity) depending on the model parameters, in particular, for some specific regions of the coupling parameter(s) of the interaction function and the dark energy equation of state. Similar to the early time instability problems in the interaction models, as discussed in section \ref{sec-dyn-analysis}, one can clearly understand that the appearance of finite time future singularities is also related to the choice of  interaction function. Since one can construct a number of phenomenological parametrizations of the 
interaction functions, thus, it is possible to have an interaction model with no finite time  singularities in the future. In this connection, we refer to an  appealing  interacting DE-DM theories where DE acts as a scalar field and the mass of the DM particles has a  direct dependence on the scalar field itself \cite{Wetterich:1987fk,Pourtsidou:2013nha}. In such theories, if the potential and kinetic term of the scalar are well behaved, no future singularity will appear.

%&&&&&&&&&&&&&&&&&&&&&&&&&&&&&&&&&&&&&&&&&&&&&&

\section{Conclusions}
\label{sec-conclu}

The theory of non-gravitational interaction between DM and DE is one of the fantastic areas of modern cosmology and this is the main theme of this work. 
Existing articles demand that 
interacting DE models are one of the promising cosmological models that could explain many theoretical and observational issues related to the evolution of the universe.  Being recognized for its ability to soften the cosmological constant problem it came into the limelight for providing with a possible solution to the cosmic coincidence problem. Now, if we concentrate on the recent investigations focused on the tensions in the cosmological parameters arising from local and global measurements, this area has taken a serious role. The readers have already witnessed how the tensions in both $H_0$ and $\sigma_8$ can be alleviated/solved if an interaction in the dark sector is considered. Therefore, there are enough reasons to select the interacting models as the theme of the present work.  In the present article we consider an interacting cosmic scenario between DM and DE driven by a simple interaction rate which is linear in the energy densities of DM and DE. Here, DM has been taken to be pressure-less and DE has a constant barotropic equation of state. By performing a systematic dynamical analysis we show explicitly that the inclusion of an interaction in the background may lead to unphysical behaviour in terms of negative energy densities. To obtain viable cosmic scenarios, one needs to impose additional conditions on the parameter space consisting of the coupling parameters of the interaction rates as well as the DE equation of state. Although the linear interaction models have been investigated widely in the past due to their simplest mathematical structure, but the current article raises some important points that are essential to understand the actual parameter space of the underlying cosmological model.

It is important to mention that the linear interaction models can also lead to finite time future cosmic singularities depending on the model parameters. In particular, we find that for the present interaction model  our universe may encounter with a BIG RIP singularity for some specific values of the model parameters. Although the present study focuses on the linear interacting models, but the same can be performed with the nonlinear interaction models. Moreover, we here focus on the simplest case in which DE has a constant barotropic state parameter, however, one could extend the case with dynamical state parameter. In connection with the present interaction model we would like to emphasize on a broad class of interacting DM-DE theories in which DE acts as a scalar field and DM particle has a mass which is directly dependent on the scalar field itself \cite{Wetterich:1987fk,Pourtsidou:2013nha}. This class of interaction theories are very appealing for their far reaching possibilities. In particular, concerning the singularity problem that we faced with the present interaction model can be avoided in the aforementioned theories if the potential and kinetic term of the scalar are well behaved.  Thus, it will be interesting to perform a dynamical system analysis of the above interaction models following the similar approach as in the present article, in order to look for viable scenarios both from theoretical and observational grounds.  Such investigations will be very enchanting and  we believe that, other investigators including us, might be interested to explore the deeper physical insights with such models.

\section*{Acknowledgments}

The authors thank the referees for their useful comments that helped us to improve the quality of discussion of this article. SP has been supported by the Mathematical Research Impact-Centric Support Scheme (MATRICS), File No. MTR/2018/000940, given by the Science and Engineering Research Board (SERB), Govt. of India.  The investigation of JdH has been supported by MINECO (Spain) grant MTM2017-84214-C2-1-P, and  in part by the Catalan Government 2017-SGR-247. WY was  supported by the  National Natural Science Foundation of China under Grants 
No. 11705079 and No. 11647153. The investigation of JA has been supported by MINECO (Spain) grant MTM2015-69135-P, and by the Catalan Government grant 2017-SGR-932.

\end{document}